\PassOptionsToPackage{hidelinks,colorlinks=true,citecolor=blue,urlcolor=blue}{hyperref}
\documentclass[amsmath, amssymb, aps, pra, onecolumn, floatfix, nofootinbib, superscriptaddress]{revtex4}

\usepackage{subfigure}
\usepackage{siunitx}
\usepackage{placeins}
\usepackage{soul}
\usepackage{braket}
\usepackage{graphicx}
\usepackage{float}
\usepackage{color}
\usepackage{booktabs}

\sethlcolor{yellow}

\newcommand{\Vscan}{V_{\mathrm{scan}}}
\newcommand{\Hstar}{H^*}
\newcommand{\Cout}{C_{\mathrm{out}}}
\newcommand{\Cin}{C_{\mathrm{in}}}
\newcommand{\Cq}{C_q}
\newcommand{\nq}{n_{\mathrm{q}}}

\begin{document}

\title{Hybrid Fourier Neural Operator for Surrogate Modeling\\ of Laser Processing with a Quantum-Circuit Mixer}

\author{Mateusz Papierz}
\affiliation{Terra Quantum AG, Kornhausstrasse 25, 9000 St.~Gallen, Switzerland}

\author{Asel Sagingalieva}
\affiliation{Terra Quantum AG, Kornhausstrasse 25, 9000 St.~Gallen, Switzerland}

\author{Alix Benoit}
\affiliation{Laboratory for Advanced Materials Processing, Feuerwerkerstrasse, Thun, 3603, BE, Switzerland}

\author{Toni Ivas}
\affiliation{Laboratory for Advanced Materials Processing, Feuerwerkerstrasse, Thun, 3603, BE, Switzerland}

\author{Elia Iseli}
\affiliation{Laboratory for Advanced Materials Processing, Feuerwerkerstrasse, Thun, 3603, BE, Switzerland}

\author{Alexey Melnikov}
\affiliation{Terra Quantum AG, Kornhausstrasse 25, 9000 St.~Gallen, Switzerland}

\begin{abstract}
Data-driven surrogates can replace expensive multiphysics solvers
for parametric PDEs, yet building compact, accurate neural operators
for three-dimensional problems remains challenging: in Fourier Neural Operators, dense mode-wise spectral channel mixing scales linearly with the number of retained Fourier modes, inflating parameter counts and limiting real-time deployability.
We introduce HQ-LP-FNO, a hybrid quantum-classical FNO that replaces a configurable fraction of these dense spectral blocks
with a compact, mode-shared variational quantum circuit mixer whose parameter count is independent of the Fourier mode budget. A parameter-matched classical bottleneck control is co-designed to provide a rigorous evaluation framework for the proposed architecture.
Evaluated on three-dimensional surrogate modeling of high-energy
laser processing, coupling heat transfer, melt-pool convection, free-surface deformation, and phase change, HQ-LP-FNO reduces
trainable parameters by 15.6\% relative to a purely classical Fourier Neural Operator baseline while lowering phase-fraction mean absolute error by 26\% and relative temperature MAE from 2.89\% to 2.56\%. A sweep over the quantum-channel budget reveals that a moderate VQC allocation yields the best temperature metrics across all tested configurations, including the fully classical baseline,
pointing toward an optimal classical-quantum partitioning. The
ablation confirms that mode-shared mixing, naturally implemented
by the VQC through its compact circuit structure, is the
dominant contributor to these improvements. A noisy-simulator study under backend-calibrated noise from \texttt{ibm\_torino} confirms numerical stability of the quantum mixer across the tested shot range.
These results demonstrate that VQC-based parameter-efficient spectral mixing can improve neural operator surrogates for complex multiphysics problems and establish a controlled evaluation protocol for hybrid quantum operator learning in physics-based settings.
\end{abstract}

\maketitle

\section{Introduction}
High-energy laser processing, including laser welding, laser powder-bed fusion, and surface remelting \cite{michaleris_modeling_2014, mukherjee_heat_2018}, enables rapid, localized modification of metallic microstructure and geometry. Predictive modeling of these processes requires resolving tightly coupled phenomena, including volumetric heat conduction, melt-pool convection, free-surface deformation, phase change, recoil-pressure-driven keyhole dynamics, and evaporation. While state-of-the-art multiphysics solvers can capture these mechanisms with high fidelity, their computational cost inhibits systematic process-window exploration, uncertainty quantification, and integration into real-time digital twins and control loops.
 
Operator learning provides a principled route to fast surrogates for parametric partial differential equation (PDE) families by learning mappings between function spaces rather than fixed-resolution pointwise predictors \cite{kovachki2023neural}. Among practical architectures, the Fourier Neural Operator (FNO) performs global mixing via FFTs and applies learnable complex-valued transformations to a truncated set of low-frequency Fourier modes \cite{li_fourier_2021}. In prior work \cite{benoit2026lpfno}, we introduced the Laser Processing Fourier Neural Operator (LP-FNO), a three-dimensional FNO surrogate trained on FLOW-3D WELD\textsuperscript{\textregistered} simulations of single-track Ti--6Al--4V processing. To make learning tractable across conduction and keyhole regimes, we reformulated the transient dynamics in a reference frame moving with the laser and applied a short sliding temporal average, yielding quasi-steady, geometry-consistent training targets. The resulting surrogate enables rapid inference of 3D temperature fields across a broad process window and supports controlled super-resolution evaluation on finer meshes. However, dense mode-wise spectral channel mixing, the mechanism by which FNO layers couple information across output channels at every retained frequency,is the dominant source of parameter growth in three dimensions, scaling linearly with the number of retained Fourier modes. This scaling limits the practicality of deploying such surrogates for real-time control or uncertainty quantification workflows, motivating the search for more compact spectral mixing strategies.
 
A natural question is therefore whether effective spectral mixing requires independent parameters at every frequency, or whether a single \emph{mode-shared} mixer, with parameters independent of the mode count, can achieve comparable fidelity at lower cost. Hybrid quantum--classical neural networks offer one such strategy: variational quantum circuits (VQCs) implement nonlinear feature maps in high-dimensional feature Hilbert spaces spaces through data-encoding gates and entanglement-induced correlations \cite{qml_review_2023, schuld2019quantummachinelearning, havlicek2019quantumfeaturespaces}, and data re-uploading architectures endow them with the ability to represent truncated Fourier series whose expressivity grows with encoding repetitions rather than parameter count \cite{schuld2021encoding, perez2020data, goto2021universal, jerbi2023quantum, kordzanganeh2023exponentially}. The quantum Fourier transform further provides a natural mechanism for spectral operations on exponentially many basis states using only logarithmically many qubits \cite{nielsen2010quantumcomputation}. Hybrid architectures embedding compact VQC layers into classical backbones have shown practical benefits in diverse domains \cite{fan2024hybrid, landman2022quantum, ajlouni2023medical, xiang2024quantum, rainjonneau2023quantum, sagingalieva2023hyperparameter, lusnig2024hybrid, anoshin2024hybrid, lee2025predictive, kurkin2025forecasting}, including PDE surrogates \cite{xiao2024piqnn, berger2025teqpinn, sedykh2024hybrid, sedykh2025multi}. However, to date no study has applied hybrid quantum classical operator learning to high-energy laser processing, a domain where coupled thermo-fluid phenomena, sharp phase interfaces, and keyhole dynamics pose particular challenges for surrogate fidelity, nor has a rigorous protocol been established to disentangle mode-shared mixing gains from any quantum-specific inductive bias.

In this work, we replace a tunable fraction of the dense spectral mixing block in LP-FNO with a compact, mode-shared VQC mixer, yielding the Hybrid Quantum LP-FNO (HQ-LP-FNO). The VQC naturally implements a structured, parameter-efficient channel interaction whose complexity is independent of the retained Fourier mode count. To provide a rigorous evaluation framework, we co-introduce a parameter-matched classical bottleneck MLP control (CM-LP-FNO), enabling precise attribution of the observed gains.

Applied to the same FLOW-3D WELD\textsuperscript{\textregistered} 
Ti--6Al--4V dataset and evaluation protocol 
of~\cite{benoit2026lpfno}, spanning conduction and keyhole welding regimes, HQ-LP-FNO reduces trainable parameters by $15.6\%$ while lowering relative temperature MAE from $2.89\%$ to $2.56\%$ and 
phase-fraction MAE by $26\%$. A noisy-simulator study under backend-calibrated noise from \texttt{ibm\_torino} confirms numerical stability of the VQC mixer, providing concrete circuit-execution benchmarks for near-term hardware deployment. Our contributions are: the first application of hybrid quantum-classical operator surrogates to high-energy laser processing, a controlled evaluation protocol for hybrid quantum spectral learning in physics-based surrogates, and a noisy-simulator validation supporting hardware viability of the proposed architecture.

\section{Methodology}\label{sec:data}
We reuse the simulation dataset, preprocessing pipeline, and train/validation/test protocol from \cite{benoit2026lpfno}. To keep the presentation self-contained while avoiding repetition, we summarize only the elements required to interpret the hybrid quantum modification introduced later, and refer readers to \cite{benoit2026lpfno} for full simulation and preprocessing details.

\subsection{High-fidelity simulation campaign and stored fields}
Ground-truth fields are generated with the commercial multiphysics solver FLOW-3D WELD 2025R1 \cite{noauthor_flow3d_weld_2025}.
The model couples incompressible flow with heat transfer and phase change, tracks free surfaces via a volume-of-fluid (VoF) formulation, and represents evaporation and keyhole recoil pressure through dedicated source terms \cite{svenungsson2015laser, aggarwal2024unravelling, flint2023laserbeamfoam}.
Laser absorption changes induced by evolving keyhole geometry are captured by ray tracing.
For each simulated process setting, we store the temperature field $T(\mathbf{x},t)$ on a uniform Cartesian grid together with the VoF metal fraction $\alpha(\mathbf{x},t)\in[0,1]$, which serves as an interface indicator and supports masking of gas-phase regions in evaluation.

The computational domain is a rectangular cuboid of dimensions
\(0.9\,\mathrm{mm} \times 0.4\,\mathrm{mm} \times 0.3\,\mathrm{mm}\)
(length \(\times\) width \(\times\) height), discretized using a uniform
mesh with a cell size of \(0.01\,\mathrm{mm}\).
Transient outputs are written every $5\,\mu\mathrm{s}$, and the total simulation time is chosen such that the laser traverses a \(0.6\,\mathrm{mm}\) scan track.

\subsection{Process window sampling via normalized enthalpy}
Laser power and scan speed span \(P \in [40\,\mathrm{W}, 190\,\mathrm{W}]\) and \(\Vscan \in [0.1\,\mathrm{m/s}, 1\,\mathrm{m/s}]\), covering lack-of-fusion (insufficient melting), conduction (surface-limited melting without a deep sustained vapor cavity), and keyhole (deep-penetration melting with a deep vapor cavity) regimes. To obtain balanced regime coverage, process settings are sampled using the normalized enthalpy criterion \(\Hstar\) \cite{hann2011laserweld},
\begin{equation}
\label{eq:Hstar}
\Hstar \;=\;
\frac{\eta\,P}
{\rho\,C_p\,\Delta T_m\,
\sqrt{\pi\,D\,\sigma^3\,\Vscan}},
\end{equation}
with parameters listed in Table~\ref{tab:nenth_params}.
Following the classical study, we sample approximately uniformly in \((\Hstar,P)\) and compute \(\Vscan\) accordingly.

\begin{figure}[!htbp]

    \centering
    \includegraphics[width=\linewidth]{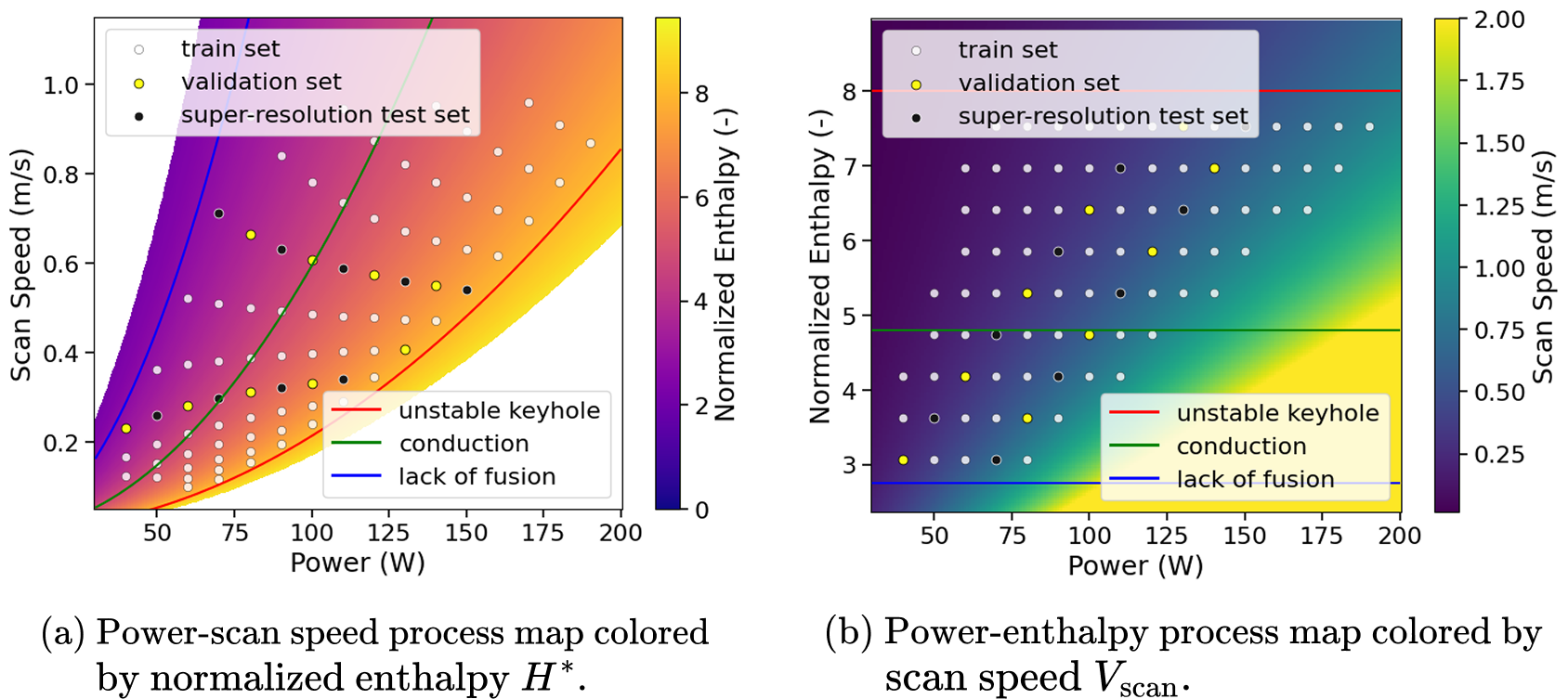}

    \caption{Process-space coverage of high-fidelity simulation data generated with FLOW-3D WELD\textsuperscript{\textregistered} \cite{benoit2026lpfno}. The dataset spans laser power \(P\) and scan speed \(\Vscan \in [0.1\,\mathrm{m/s}, 1.0\,\mathrm{m/s}]\), and is constructed on a grid uniform in normalized enthalpy \(\Hstar\) and power. Markers denote the training, validation, and super-resolution test sets used for neural operator learning, (Figure adapted from~\cite{benoit2026lpfno}).}
    \label{fig:enthalpy_plots}
\end{figure}

Figure~\ref{fig:enthalpy_plots} provides a compact overview of the process-space coverage and the split into training, validation, and super-resolution test sets reused throughout this follow-up. Visualizing the dataset in both the \((P,\Vscan)\) and \((P,\Hstar)\) planes highlights that the sampling is approximately uniform in normalized enthalpy while spanning the full power range, thereby ensuring balanced representation of these three regimes under a single, consistent protocol.

\begin{table}[!htbp]
\centering
\caption{Parameters used for normalized enthalpy evaluation (same as the classical study).}
\label{tab:nenth_params}
\begin{tabular}{lll}
\toprule
Symbol & Value & Description \\
\midrule
$\eta$ & \SI{0.35}{} & Absorptivity \\
$\rho$ & \SI{4420}{\kg\per \m^{3}} & Density \\
$C_p$ & \SI{750}{\J\per \kg \per \K} & Specific heat capacity at Tsolidus=1873K \\
$\Delta T_m$ & \SI{1573}{K} & Temperature rise to melting \\
$D$ & \SI{8.1e-6}{\m^2/\s} & Thermal diffusivity \\
$\sigma$ & \SI{50}{\micro\meter} & Beam radius (1/e$^2$) \\
$L$ & \SI{3.45e5}{\J \per \kg} & Latent Heat of Fusion \\
$P$ & \SI{40}{}--\SI{190}{\W} & Laser power range \\
$V_{\text{scan}}$ & \SI{0.1}{}--\SI{1}{\m/\s} & Scan speed range \\
\bottomrule
\end{tabular}
\end{table}

\subsection{Quasi-steady moving-frame representation}
Operator learning is performed on quasi-steady fields in a reference frame moving with the laser at constant speed $\Vscan$.
In preprocessing, we subsample the stored transient outputs so that the retained increment satisfies
\begin{equation}
\Delta t = \frac{\Delta x}{\Vscan},
\end{equation}
which avoids spatial interpolation on the uniform mesh.
Residual temporal variation in the moving frame is dominated by high-frequency interface fluctuations (especially in keyhole regimes), we therefore apply a sliding temporal average over a 30-step window to isolate the large-scale melt-pool structure, consistent with the classical preprocessing.

\subsection{Learning task, derived indicators, and normalization}
The learned surrogate approximates a parametric solution operator that maps process parameters to the three-dimensional temperature field and VoF metal-fraction field on the spatial grid in the moving laser frame. Consistent with standard FNO implementations and with \cite{benoit2026lpfno}, we augment constant process parameters with coordinate features so that each grid point receives the input tuple \((x,y,z,P,\Vscan, H^*)\), and we train the network to predict the fields \(T(\mathbf{x})\) and \(\alpha(\mathbf{x})\). The VoF metal fraction \(\alpha(\mathbf{x}) \in [0,1]\) is thus treated as a direct learned output alongside temperature and is used in this work for masking and for defining metal-region quantities of interest.

To evaluate melt-related quantities derived from temperature, we compute a liquid fraction proxy $f_l$ from the (predicted and reference) temperature using a piecewise-linear rule between Ti--6Al--4V solidus and liquidus temperatures \cite{Boivineau2006Ti64Thermo}.

\begin{table}[!htbp]
\centering
\caption{Reference scales used for normalization (SI units), same as the classical study.}
\label{tab:reference_scales}
\begin{tabular}{lll}
\toprule
Quantity & Symbol & Reference value \\
\midrule
Length scale        & $L$    & \SI{1e-4}{\meter} \\
Temperature scale   & $T$    & \SI{3000}{\kelvin} \\
Velocity scale      & $V$    & \SI{1e-1}{\meter\per\second} \\
Power scale         & $P$    & \SI{10}{\watt} \\
Normalized enthalpy & $H$    & \num{7.5} \\
\bottomrule
\end{tabular}
\end{table}

\paragraph{Gas-phase masking.}
Quantities of interest are confined to the metal region. For evaluation we therefore apply a smooth VoF-based mask to suppress artifacts in the surrounding gas phase. Throughout, the mask is constructed from the reference $\alpha$ provided by the solver.

Let $\alpha(\mathbf{x})$ denote the metal volume fraction and define
\begin{equation}
g(\mathbf{x}) = \frac{1}{2}\left[\tanh\!\left(k(\alpha(\mathbf{x}) - 0.5)\right) + 1\right],
\end{equation}
where $k$ controls the sharpness of the transition (we use $k=20$ as in the classical study).

We blend the temperature toward a fixed high-temperature value $T_{\mathrm{boil}}$ in the gas region,
\begin{equation}
\tilde{T}(\mathbf{x}) = T_{\mathrm{boil}} + g(\mathbf{x})\bigl(T(\mathbf{x}) - T_{\mathrm{boil}}\bigr),
\end{equation}
with $T_{\mathrm{boil}}=\SI{3123}{K}$ \cite{Zhang2020Ti64Boiling} (simulation setting), and we mask the liquid fraction to zero outside the metal.
Reported temperature errors are computed on the masked field $\tilde{T}$ to avoid spurious gas-phase contributions.

\paragraph{Evaluation metrics and reporting.}
Following LP-FNO \cite{benoit2026lpfno}, we evaluate field-level fidelity using mean absolute error (MAE) and root-mean-square error (RMSE) on the masked temperature field $\tilde{T}$ and on the melt-related fields $\alpha$ and $f_l$. For temperature we additionally report relative metrics computed using a global normalization constant $\mu_y$ (the mean of the ground-truth values over the evaluation set),
\begin{equation}
\mu_y \;=\; \frac{1}{N\,|D_h|}\sum_{s=1}^{N}\sum_{\mathbf{x}\in D_h} y^{(s)}(\mathbf{x}),
\end{equation}
and
\begin{equation}
\mathrm{RelMAE} \;=\; \frac{\mathrm{MAE}}{\mu_y + \varepsilon},
\qquad
\mathrm{RelRMSE} \;=\; \frac{\mathrm{RMSE}}{\mu_y + \varepsilon},
\end{equation}
with a small $\varepsilon>0$ for numerical stability. For $\alpha$ and $f_l$ we additionally report the intersection over union (IoU) of thresholded masks with $\tau=0.5$. Absolute temperature errors are reported in Kelvin by rescaling the normalized temperature output with the same reference scale $T_{\mathrm{ref}}=\SI{3000}{K}$ used during training; $\alpha$ and $f_l$ are already dimensionless in $[0,1]$.

Since the present work targets parameter efficiency rather than runtime, we interpret performance through a Pareto lens: a hybrid model is considered beneficial if it matches or improves these errors while reducing the dominant spectral parameter count, and crucially if it does so without a loss in generalization under the held-out process-window and super-resolution evaluations reused from the classical baseline.

\subsection{Ablation design and evaluation protocol}
\label{sec:ablation_design}
 
To isolate the effect of mode-shared spectral mixing from any quantum-specific inductive bias, we design a controlled ablation around a single integer knob, the \emph{quantum-channel width} $\Cq \in \{0,1,\dots,C\}$, which determines how many of the $C$ spectral output channels per retained Fourier mode are produced by a compact, mode-shared mixer rather than by the standard dense, mode-wise complex spectral weights. Setting $\Cq=0$ recovers the classical LP-FNO baseline of \cite{benoit2026lpfno}, increasing $\Cq$ removes dense spectral parameters that scale with the number of retained Fourier modes $N_{\mathrm{m}}$ and replaces them with a shared mixer whose parameter count is independent of $N_{\mathrm{m}}$ (explicit accounting is given in Sec.~\ref{sec:params}).
 
For each value of $\Cq$ we train two mode-shared mixer variants under otherwise identical conditions:
\begin{itemize}
    \item \textbf{HQ-LP-FNO}: the shared mixer is a variational quantum circuit (VQC) embedded in a QFT--mixer--inverse-QFT structure (Sec.~\ref{sec:hybrid_lp_fno});
    \item \textbf{CM-LP-FNO}: the shared mixer is a purely classical bottleneck MLP whose trainable-parameter budget is matched to that of the VQC at the same $\Cq$.
\end{itemize}
Because HQ-LP-FNO and CM-LP-FNO share the same channel partitioning, spectral routing, and total parameter budget, any accuracy difference between them can be attributed to the mixer realization (quantum vs.\ classical) rather than to model capacity or data protocol.
 
We quantify the resulting parameter--accuracy trade-off via a sweep over $\Cq$ (Sec.~\ref{sec:results_ratio_sweep}), with the classical LP-FNO ($\Cq=0$) as the reference. Neither the data generation, preprocessing, train/validation/test splits, nor the evaluation metrics and masking protocol of \cite{benoit2026lpfno} are changed; only the spectral channel-mixing parameterization inside the Fourier layers differs between runs. To assess robustness beyond idealized noiseless simulation, we additionally validate the quantum mixer under backend-calibrated noise from \texttt{ibm\_torino}, confirming numerical stability across a practical shot range (Appendix~\ref{app:noisy_simulator}).

\section{Hybrid Quantum LP-FNO}
\label{sec:hybrid_lp_fno}

\subsection{LP-FNO baseline}
\label{sec:classical_fno}
We take the LP-FNO architecture introduced in \cite{benoit2026lpfno} as the classical baseline and keep its architecture, preprocessing, data splits, optimization, and evaluation protocol fixed. Consequently, any performance or parameter-count differences reported later can be attributed solely to the modification introduced here: a different parameterization of low-frequency spectral channel mixing inside each Fourier layer.

Let $a(\mathbf{x})\in\mathbb{R}^{\Cin}$ denote the input feature field on a regular 3D grid (here: spatial coordinates together with constant process parameters) and let $u_\ell(\mathbf{x})\in\mathbb{R}^{C}$ be the latent representation at layer $\ell$. An FNO lifts $u_0=P(a)$, applies $L$ Fourier layers, and projects by $Q$:
\begin{equation}
u_{\ell+1}(\mathbf{x})
=
\sigma\!\left(
W_\ell u_\ell(\mathbf{x}) +
\mathcal{F}^{-1}\!\left(\mathcal{K}_\ell\big(\hat{u}_\ell\big)\right)(\mathbf{x})
\right),
\qquad
\hat{u}_\ell=\mathcal{F}[u_\ell],
\label{eq:fno_layer_classical}
\end{equation}
where $W_\ell$ is a pointwise linear map ($1\times 1\times 1$ convolution), $\sigma$ is GELU \cite{hendrycks2016gelu} (applied after each Fourier layer except the last, as in our implementation), and $\mathcal{F}$ denotes the real 3D FFT (\texttt{rfftn}). The learnable spectral operator $\mathcal{K}_\ell$ acts only on a truncated low-frequency index set $\Omega$, coefficients outside $\Omega$ are set to zero, and the inverse FFT returns a spatial field with global, low-frequency coupling \cite{li_fourier_2021, kovachki2023neural}. Because the mapping is defined in Fourier space, the trained operator can be evaluated at grid resolutions different from the training resolution (super-resolution testing).

For each retained mode $k\in\Omega$, write $\hat{u}_\ell(k)\in\mathbb{C}^{C}$ (complex Fourier coefficients across channels). The baseline spectral update is a dense \emph{mode-wise} complex linear map,
\begin{equation}
\mathcal{K}_\ell\big(\hat{u}_\ell\big)(k)=\hat{u}_\ell(k)\,R_\ell(k),
\qquad
R_\ell(k)\in\mathbb{C}^{C\times C},\quad k\in\Omega,
\end{equation}
implemented in practice by storing four complex weight tensors (``four corners'' in $(k_x,k_y)$ combined with one-sided frequencies along the rFFT axis) \cite{li_fourier_2021, kovachki2023neural}. This dense per-mode channel mixing is precisely the component we modify in the hybrid construction below.

For controlled comparison, we keep the backbone architecture, data protocol, and evaluation pipeline aligned with \cite{benoit2026lpfno}. The hybrid model differs from the baseline in the spectral channel-mixing parameterization within $\Omega$ and, in training, uses a cosine-annealing learning-rate schedule with a higher base learning rate (Table~\ref{tab:fno_hparams}), reflecting the step-size sensitivity commonly observed in variational quantum optimization \cite{koelle2024vqeoptimization}.

\subsection{Partitioned hybrid architecture}

We build on the partitioned hybridization principle of PH-QFNO \cite{Marcandelli2025HybridFNO} and Fourier-based quantum operator learning \cite{Jain2024QFNO}, tailoring it to our 3D laser--material surrogate setting. Our objective is \emph{parameter-efficient spectral learning} under a strictly controlled protocol: we keep the classical 3D FFT backbone and the full FNO computation graph intact, and replace only a configurable fraction of low-frequency \emph{spectral channel-mixing outputs} by features generated through a compact mode-shared mixer. We consider two instantiations: a VQC-based mixer (HQ-LP-FNO) and a parameter-matched classical bottleneck MLP (CM-LP-FNO). Because the quantum block is simulated in this work, we do not claim runtime acceleration. Instead, we isolate representational and parameterization effects by benchmarking accuracy and generalization under the same held-out and super-resolution protocols as the classical baseline.

\subsubsection{Partitioned hybrid spectral convolution}
We modify only the learnable spectral operator $\mathcal{K}_\ell$ in Eq.~\eqref{eq:fno_layer_classical}. Let $\Omega$ denote the retained low-frequency index set in the \texttt{rfftn} representation (standard four-corner truncation in $(k_x,k_y)$ and one-sided frequencies along the rFFT axis) \cite{li_fourier_2021, kovachki2023neural}. Using an integer \emph{quantum-channel width} $\Cq\in\{0,1,\ldots,C\}$, we split the \emph{output} channels into a quantum subset and a classical subset \cite{kordzanganeh2023parallel},
\begin{equation}
\Cq\in\{0,1,\ldots,C\},\qquad C_c=C-\Cq.
\label{eq:channel_split}
\end{equation}
Since all Fourier layers in our architecture share the same channel width $C$ (set by the lifting projection $P$), the constraint simplifies to $\Cq \le C$.

With this convention, $\Cq$ does not change the spatial-frequency truncation: both branches operate on the same retained mode set $\Omega$ (the standard four-corner truncation described below). Instead, $\Cq$ partitions the \emph{channel dimension} at every retained mode. Concretely, for each $k\in\Omega$ we route the first $\Cq$ complex channels $\hat{u}^{(q)}_\ell(k)$ to the quantum mixer and generate the first $\Cq$ output channels $\hat{v}^{(q)}_\ell(k)$, while the remaining $C_c$ output channels are produced by classical mode-wise mixing. In our implementation we set the circuit width to the quantum-channel width, $\nq=\Cq$ (unless stated otherwise), so increasing $\Cq$ increases both the number of quantum-generated channels per mode and the number of qubits. Operationally, the VQC is applied to each retained mode instance as a $2\Cq$-dimensional real vector obtained by concatenating real and imaginary parts; thus, for batch size $B$, one forward pass processes $B\times 4 m_x m_y m_z$ such instances, with the same circuit parameters shared across all $k\in\Omega$.

For each retained Fourier mode $k\in\Omega$, write $\hat{u}_\ell(k)\in\mathbb{C}^{C}$. The hybrid spectral update produces $\hat{v}_\ell(k)\in\mathbb{C}^{C}$ by concatenating a quantum and a classical branch,
\begin{equation}
\hat{v}_\ell(k)
=
\begin{bmatrix}
\hat{v}^{(q)}_\ell(k)\\[2pt]
\hat{v}^{(c)}_\ell(k)
\end{bmatrix}
=
\begin{bmatrix}
\mathcal{Q}_\theta\!\big(\hat{u}^{(q)}_\ell(k)\big)\\[2pt]
\hat{u}_\ell(k)\,R^{(c)}_\ell(k)
\end{bmatrix},
\qquad k\in\Omega,
\label{eq:partitioned_update}
\end{equation}
and $\hat{v}_\ell(k)=0$ for $k\notin\Omega$. Here $\hat{u}^{(q)}_\ell(k)\in\mathbb{C}^{\Cq}$ denotes the first $\Cq$ input channels at mode $k$, $R^{(c)}_\ell(k)\in\mathbb{C}^{C\times C_c}$ are classical complex spectral weights producing the remaining $C_c$ output channels, and $\mathcal{Q}_\theta$ is a \emph{mode-shared} quantum map producing the first $\Cq$ output channels. Importantly, the partitioning changes only the channel-mixing parameterization inside $\Omega$: the classical branch remains \emph{mode-wise} (distinct complex weights per retained mode), while the quantum branch applies the same circuit parameters to every processed mode (shared across all $k\in\Omega$ and across samples).

In practice, our implementation follows the standard four-corner layout in $(k_x,k_y)$ and the one-sided rFFT axis. For numerical robustness under resolution changes, the effective number of retained modes is computed in the forward pass as
$m_x=\min(m_x^{\text{set}},\lfloor X/2\rfloor)$,
$m_y=\min(m_y^{\text{set}},\lfloor Y/2\rfloor)$,
$m_z=\min(m_z^{\text{set}},Z_{\mathrm{fft}})$,
where $Z_{\mathrm{fft}}$ is the rFFT size along the last axis. This ensures that the hybrid spectral layer is well-defined for both the training grid and super-resolution evaluation grids.

\paragraph{Parameter-matched classical mixer control, CM-LP-FNO.}
To disentangle quantum-specific inductive bias from the structural effect of replacing dense mode-wise mixing by a mode-shared mixer, we introduce a purely classical control variant, CM-LP-FNO. This control keeps the same channel partitioning and spectral routing as HQ-LP-FNO but replaces the VQC with a mode-shared bottleneck MLP applied independently to each retained Fourier mode, acting on the concatenated real and imaginary parts of the \(\Cq\) complex Fourier coefficients. In the default setting, the same input scaler as in the quantum branch is applied first, after which the mixer uses a linear input projection, GELU activation, optional additional hidden linear layers for depth \(>1\), and a linear output projection back to \(2\Cq\) real features, which are then reassembled into complex coefficients. The bottleneck width is chosen to approximately match the trainable-parameter budget of the VQC at fixed \(\Cq\); all other architecture and training settings are identical to the HQ-LP-FNO runs.

\begin{figure}[!htbp]
    \centering
    \includegraphics[width=\linewidth,height=0.75\textheight,keepaspectratio]{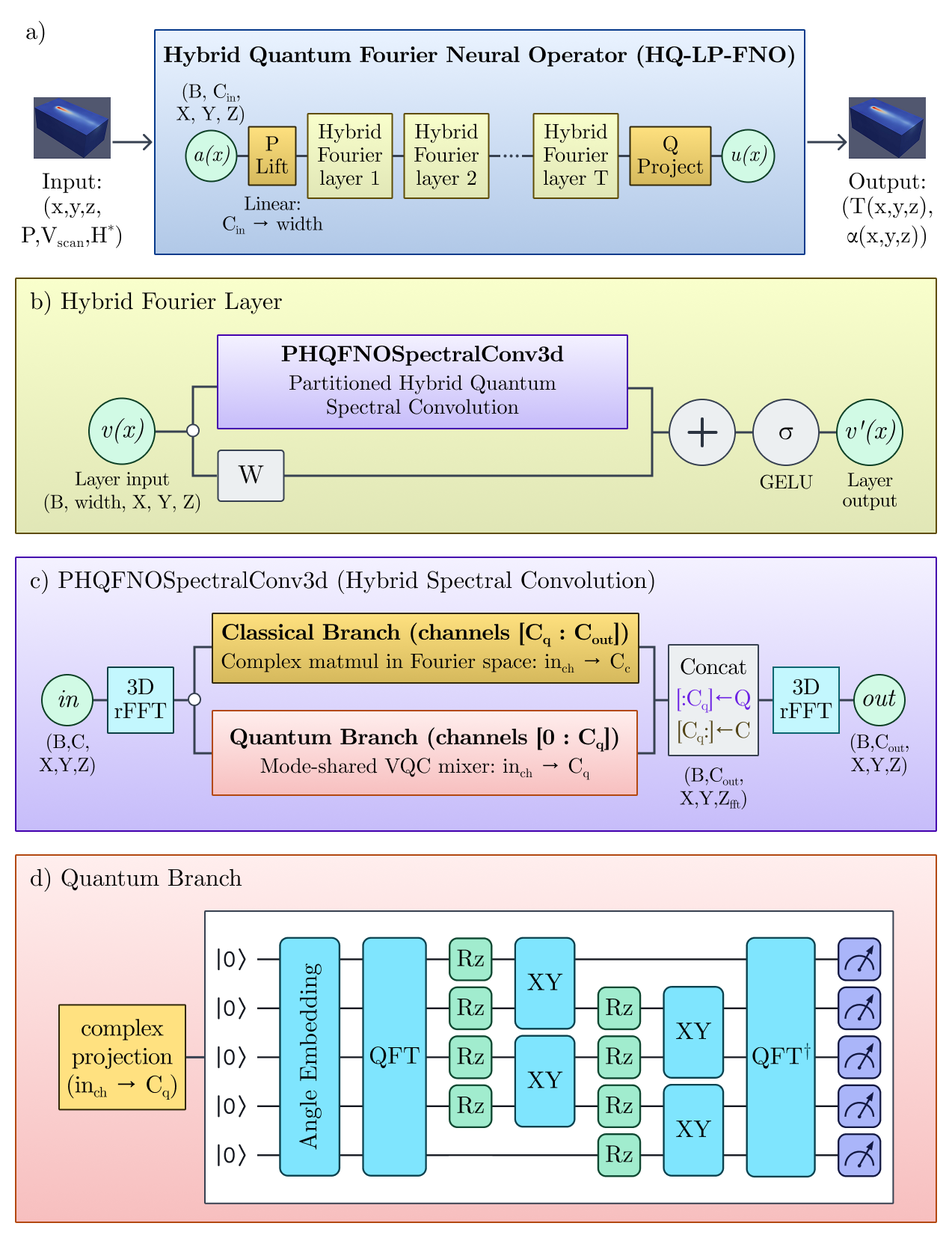}
    \caption{HQ-LP-FNO for 3D laser--material heat transfer.
    (a) Input features are lifted by a pointwise projection $P$, processed by $L$ hybrid Fourier layers, and projected back by $Q$ to predict $T$ and $\alpha$.
    (b) One hybrid Fourier layer: a spectral path (PHQFNOSpectralConv3d) runs in parallel with a pointwise linear map $W_\ell$, outputs are summed and passed through GELU (except in the last Fourier layer, where no nonlinearity is applied).
    (c) Partitioned hybrid spectral convolution: after a 3D real FFT, output channels are split into a classical branch (channels $[\Cq:\Cout]$) with learnable complex spectral weights, and a quantum branch (channels $[0:\Cq]$) that applies a mode-shared variational quantum circuit to the first $\Cq$ input channels. The two branches are concatenated and mapped back via an inverse 3D real FFT.}
    \label{fig:PH-QFNO}
\end{figure}

\subsubsection{Quantum spectral mixer block}
The quantum branch operates on complex Fourier coefficients over the first $\Cq$ channels. For each mode $k\in\Omega$, we construct a real feature vector by concatenating real and imaginary parts,
\begin{equation}
r_\ell(k)=\big[\Re\,\hat{u}^{(q)}_\ell(k)\;\; \Im\,\hat{u}^{(q)}_\ell(k)\big]\in\mathbb{R}^{2\Cq}.
\end{equation}
Because direct mapping of Fourier coefficients to rotation angles can induce phase wrapping and unstable optimization, we apply a robust, feature-wise percentile min--max scaler with exponential moving averages: during training, effective minima/maxima are estimated as the 2nd and 98th percentiles per feature and tracked with momentum $\mu=0.05$, while at inference the stored running estimates are reused. Values outside the percentile range are softly compressed rather than hard-clipped to preserve gradients. Concretely, letting $q_{0.02}(\cdot)$ and $q_{0.98}(\cdot)$ denote feature-wise percentiles over the current batch, we update
\begin{equation}
m \leftarrow (1-\mu)\,m + \mu\,q_{0.02}(r),\qquad
M \leftarrow (1-\mu)\,M + \mu\,q_{0.98}(r),
\label{eq:ema_percentiles}
\end{equation}
then normalize, apply a centered sigmoid, and scale to angles in $(0,\pi)$,
\begin{equation}
r_{\mathrm{norm}}=\frac{r-m}{(M-m)+\varepsilon},\qquad
s=\sigma\!\big(6(r_{\mathrm{norm}}-0.5)\big)\in(0,1),\qquad
\tilde{r} = \pi\, s \in (0,\pi)^{2\Cq}.
\label{eq:robust_scaling}
\end{equation}

We then apply learned linear pre-/post-projections with bias, $E:\mathbb{R}^{2\Cq}\!\to\mathbb{R}^{\nq}$ and $D:\mathbb{R}^{\nq}\!\to\mathbb{R}^{2\Cq}$, and set $\nq=\Cq$ (one qubit per quantum channel) in this work. The circuit follows a QFT--mixer--inverse-QFT structure. After \texttt{AngleEmbedding} of $E\tilde{r}_\ell(k)$ (default $R_X$ rotations), we apply a QFT across the register, a depth-$d$ nearest-neighbour odd--even IsingXY mesh with local $R_Z$ phases, and the inverse QFT:
\begin{equation}
\ket{\psi_\theta(k)} =
\mathrm{QFT}^\dagger\;
U_\theta\;
\mathrm{QFT}\;
\left(\bigotimes_{j=1}^{\nq} R_X(\alpha_{\ell,j}(k))\right)\ket{0}^{\otimes \nq},
\end{equation}
where QFT is the standard unitary \cite{nielsen2010quantumcomputation}. The variational mixer $U_\theta$ is implemented as a depth-$d$ nearest-neighbour chain with odd--even pairing; across one depth layer it covers all adjacent pairs once, resulting in $(\nq-1)$ pair operations per layer. Each pair operation carries two local $R_Z$ phases and one exchange-type coupling, implemented using the two-qubit \texttt{IsingXY} gate \cite{abrams2020xyfamily},
\begin{equation}
\mathrm{XY}(\phi)=\exp\!\left(i\frac{\phi}{4}(X\!\otimes\!X+Y\!\otimes\!Y)\right)
=
\begin{pmatrix}
1 & 0 & 0 & 0\\
0 & \cos(\phi/2) & i\sin(\phi/2) & 0\\
0 & i\sin(\phi/2) & \cos(\phi/2) & 0\\
0 & 0 & 0 & 1
\end{pmatrix},
\end{equation}
which acts nontrivially only on the $\{|01\rangle,|10\rangle\}$ subspace. Finally, we measure Pauli-$Z$ expectation values on each qubit, decode them via $D$ to obtain $2\Cq$ real numbers, and reshape these into $\hat{v}^{(q)}_\ell(k)\in\mathbb{C}^{\Cq}$, which populates the first $\Cq$ output channels at mode $k$ in Eq.~\eqref{eq:partitioned_update}. In implementation, retained modes are treated as independent batch items for the quantum mixer (per corner block) by reshaping $(B,\Cq,m_x,m_y,m_z)$ into a batch of size $B\,m_xm_ym_z$ via \texttt{view\_as\_real}, applying the shared quantum block, reshaping back via \texttt{view\_as\_complex}, and scattering to the corresponding Fourier corner. The circuit is simulated and differentiated end-to-end using PennyLane's \texttt{default.qubit} backend with backpropagation \cite{bergholm2018pennylane, kuzmin2025tqml}.

\subsection{Parameter accounting}
\label{sec:params}

We provide explicit parameter accounting for the spectral branch, which is the dominant source of parameter growth in 3D FNO layers due to its dependence on the number of retained Fourier modes. Pointwise parameters ($P$, $Q$, and $W_\ell$) are identical between the classical and hybrid models and are therefore omitted from the comparison; any parameter differences arise exclusively from replacing a fraction of the mode-wise complex spectral weights by a mode-shared quantum mixer with learned classical pre-/post-projections.

\paragraph{Classical spectral parameters, 3D FNO.}
Consider one 3D spectral convolution at channel width $C$. Let $(m_x,m_y,m_z)$ denote the effective retained-mode counts used in the forward pass (with $m_x\le\lfloor X/2\rfloor$, $m_y\le\lfloor Y/2\rfloor$, and $m_z\le Z_{\mathrm{fft}}$ for an input grid of size $X\times Y\times Z$ and rFFT size $Z_{\mathrm{fft}}=\lfloor Z/2\rfloor+1$). Define $N_{\mathrm{m}}=m_xm_ym_z$. Standard \texttt{rfftn}-based FNO implementations instantiate four complex weight tensors corresponding to the four $(k_x,k_y)$ corner blocks (with one-sided frequencies along the rFFT axis) \cite{li_fourier_2021, kovachki2023neural}. Each corner block contains $C^2\,N_{\mathrm{m}}$ complex parameters, i.e.\ $2\,C^2\,N_{\mathrm{m}}$ real parameters (real and imaginary parts). Hence, per spectral layer,
\begin{equation}
N_{\mathrm{spec}}^{\mathrm{FNO}}
= 4 \times 2\,C^2\,N_{\mathrm{m}}
= 8\,C^2\,N_{\mathrm{m}}.
\label{eq:param_fno}
\end{equation}

\paragraph{Hybrid spectral reduction.}
In the partitioned layer, dense classical spectral weights are learned only for the $C_c=C-\Cq$ classical output channels (Eq.~\ref{eq:channel_split}); the first $\Cq$ output channels are produced by the mode-shared quantum mixer. Therefore, the remaining mode-wise spectral parameters per layer are
\begin{equation}
N_{\mathrm{spec}}^{\mathrm{hyb}} = 8\,C\,(C-\Cq)\,N_{\mathrm{m}},
\qquad
\Delta N_{\mathrm{spec}}
= N_{\mathrm{spec}}^{\mathrm{FNO}}-N_{\mathrm{spec}}^{\mathrm{hyb}}
= 8\,C\,\Cq\,N_{\mathrm{m}}.
\label{eq:param_hyb}
\end{equation}
Thus, the knob $\Cq\le C$ controls the fraction of dense mode-wise spectral parameters removed, while leaving the classical FFT backbone unchanged.

\paragraph{Quantum-branch parameters, mode-shared, trainable.}
The quantum branch introduces trainable parameters that do \emph{not} scale with $N_{\mathrm{m}}$ because they are shared across all modes within a layer: circuit parameters of the odd--even IsingXY mesh, and the linear pre-/post-projections around the circuit. With $\nq$ qubits and depth $d$, our circuit applies $(\nq-1)$ neighbour-pair operations per depth layer, each carrying 3 angles ($R_Z$ phase, \texttt{IsingXY} angle, $R_Z$ phase), giving
\begin{equation}
N_{\mathrm{circ}} = 3\,d\,(\nq-1).
\label{eq:param_circ}
\end{equation}
The projections are standard fully connected layers with bias, $E\in\mathbb{R}^{\nq\times 2\Cq}$ and $D\in\mathbb{R}^{2\Cq\times \nq}$, hence
\begin{equation}
N_{\mathrm{proj}}
= (2\Cq\,\nq + \nq) \;+\; (2\Cq\,\nq + 2\Cq)
= 4\Cq\,\nq + \nq + 2\Cq.
\label{eq:param_proj}
\end{equation}
The robust scaler maintains running percentile statistics as non-trainable state and contributes no trainable parameters. Overall, the trainable quantum-branch parameter count per layer is
\begin{equation}
N_{q}=N_{\mathrm{circ}}+N_{\mathrm{proj}}
= 3\,d\,(\nq-1) + 4\Cq\,\nq + \nq + 2\Cq.
\label{eq:param_quantum_total}
\end{equation}

\paragraph{Per-layer and network-level scaling.}
Per layer, the hybridization removes $\Delta N_{\mathrm{spec}}$ dense mode-wise parameters (Eq.~\eqref{eq:param_hyb}) and adds $N_q$ mode-shared parameters (Eq.~\eqref{eq:param_quantum_total}). Since $\Delta N_{\mathrm{spec}}$ grows linearly with $N_{\mathrm{m}}$ while $N_q$ is independent of $N_{\mathrm{m}}$, the parameter savings are most pronounced in three-dimensional settings with large retained-mode budgets. For a network with $L$ Fourier layers, the spectral-branch totals scale approximately as $L\,N_{\mathrm{spec}}^{\mathrm{FNO}}$ for the classical baseline and $L\,(N_{\mathrm{spec}}^{\mathrm{hyb}}+N_q)$ for the hybrid model (with the same pointwise parameters in both cases).

\begin{table}[!htbp]
\centering
\caption{Trainable spectral-branch parameters per Fourier layer.}
\label{tab:param_summary}
\begin{tabular}{lll}
\toprule
Component & Classical FNO & Hybrid FNO \\
\midrule
Mode-wise spectral weights & $8\,C^2\,N_{\mathrm{m}}$ & $8\,C(C-\Cq)\,N_{\mathrm{m}}$ \\
Mode-shared quantum branch & -- & $3\,d(\nq-1)+4\Cq\nq+\nq+2\Cq$ \\
\bottomrule
\end{tabular}
\end{table}

\section{Results and Discussion}
\label{sec:results}

\subsection{Experimental Setup}
This paper is a controlled methodological follow-up to LP-FNO \cite{benoit2026lpfno}. We keep the dataset, preprocessing, splits (including held-out windows and super-resolution), and backbone architecture fixed, and modify only the low-frequency spectral channel mixing inside the Fourier layers (Sec.~\ref{sec:hybrid_lp_fno}). The training setup is matched to the baseline protocol unless stated otherwise, in particular, for the hybrid runs reported here we use cosine-annealing learning-rate scheduling ($T_{\max}=6000$, $\eta_{\min}=\num{1e-5}$) and a base learning rate of $\eta_0=\num{1e-4}$.

Unless stated otherwise, the hybrid model refers to the configuration with $\Cq=5$, which uses $\nq=\Cq$ and circuit depth $d=1$ (Sec.~\ref{sec:hybrid_lp_fno}). Temperature metrics are reported on the masked field $\tilde{T}$ as in the baseline protocol, and melt-related indicators are evaluated consistently with the classical study (Sec.~\ref{sec:data}). Absolute temperature errors are therefore reported in Kelvin by rescaling with $T_{\mathrm{ref}}=\SI{3000}{K}$. Unless stated otherwise, we denote the melt fraction by $f_l$ and use the same segmentation and overlap metrics.

\begin{table}[!htbp]
\centering
\caption{Error metrics on the held-out test set (mean $[\min,\max]$ across 8 folds, $\Cq=5$).
CM-LP-FNO replaces the VQC with a parameter-matched mode-shared bottleneck MLP.
Absolute temperature errors are reported in kelvin (rescaled with $T_{\mathrm{ref}}=\SI{3000}{K}$), other fields are dimensionless.
Best mean value per metric is shown in \textbf{bold} (lower is better for error metrics, higher is better for IoU).}
\label{tab:test_metrics_Cq5}
\renewcommand{\arraystretch}{1.15}
\setlength{\tabcolsep}{2pt}
\small
\begin{tabular}{@{}l ccc@{}}
\toprule
& \textbf{LP-FNO} & \textbf{HQ-LP-FNO} & \textbf{CM-LP-FNO} \\
& (classical) & (VQC, $\Cq\!=\!5$) & (MLP, $\Cq\!=\!5$) \\
\midrule
\multicolumn{4}{@{}l}{\textit{Temperature field $\tilde{T}$}} \\[2pt]
Abs.\ mean [K]
  & 18.2\,[14.1,\,21.4]
  & \textbf{15.2}\,[14.0,\,15.8]
  & 15.4\,[13.2,\,17.5] \\
RMSE [K]
  & 37.3\,[31.4,\,45.3]
  & 39.7\,[38.4,\,40.7]
  & \textbf{32.1}\,[28.6,\,33.8] \\
Rel.\ mean [\%]
  & 2.9\,[2.2,\,3.4]
  & 2.6\,[2.4,\,2.7]
  & \textbf{2.3}\,[2.0,\,2.6] \\
Rel.\ RMSE [\%]
  & 5.9\,[5.0,\,7.2]
  & 6.7\,[6.5,\,6.9]
  & \textbf{4.7}\,[4.2,\,5.0] \\
\midrule
\multicolumn{4}{@{}l}{\textit{Phase fraction $\alpha$}} \\[2pt]
Abs.\ mean
  & 0.0035\,[0.0025,\,0.0048]
  & \textbf{0.0026}\,[0.0024,\,0.0029]
  & \textbf{0.0026}\,[0.0022,\,0.0032] \\
RMSE
  & 0.0174\,[0.0135,\,0.0251]
  & 0.0232\,[0.0221,\,0.0237]
  & \textbf{0.0134}\,[0.0126,\,0.0145] \\
IoU mean
  & 0.9992\,[0.9986,\,0.9994]
  & 0.9988\,[0.9988,\,0.9989]
  & \textbf{0.9995}\,[0.9995,\,0.9996] \\
IoU std
  & 0.00049\,[0.00024,\,0.00106]
  & 0.00070\,[0.00061,\,0.00075]
  & \textbf{0.00034}\,[0.00023,\,0.00064] \\
\midrule
\multicolumn{4}{@{}l}{\textit{Liquid fraction $f_l$}} \\[2pt]
Abs.\ mean
  & 0.0024\,[0.0018,\,0.0034]
  & 0.0032\,[0.0028,\,0.0036]
  & \textbf{0.0012}\,[0.0010,\,0.0015] \\
RMSE
  & 0.0369\,[0.0297,\,0.0452]
  & 0.0469\,[0.0449,\,0.0490]
  & \textbf{0.0216}\,[0.0192,\,0.0273] \\
IoU mean
  & 0.9106\,[0.8771,\,0.9321]
  & 0.8653\,[0.8535,\,0.8795]
  & \textbf{0.9702}\,[0.9645,\,0.9754] \\
IoU std
  & 0.053\,[0.02,\,0.13]
  & 0.043\,[0.031,\,0.056]
  & \textbf{0.017}\,[0.00731,\,0.032] \\
\bottomrule
\end{tabular}
\end{table}

\subsection{LP-FNO vs mode-shared mixer variants}
\label{sec:results_r015}
We begin with a direct comparison between the classical LP-FNO baseline and two $\Cq=5$ variants that replace a fraction of dense mode-wise spectral channel mixing by a mode-shared mixer (Sec.~\ref{sec:hybrid_lp_fno}): the VQC-based HQ-LP-FNO and a parameter-matched purely classical control (CM-LP-FNO).

\paragraph{Parameter budget.}
The classical model has $184{,}326{,}722$ trainable parameters. At $\Cq=5$, the VQC-based HQ-LP-FNO variant has $155{,}527{,}103$ trainable parameters ($15.6\%$ fewer), consistent with the accounting in Sec.~\ref{sec:params}. The parameter-matched classical control (CM-LP-FNO) has $155{,}527{,}112$ trainable parameters (within $<10^{-5}$ relative difference to the VQC variant), isolating mixer design effects from parameter-budget effects. Both variants target the dominant parameter-growth pathway in 3D FNOs dense mode-wise channel mixing, and replace it with a mode-shared mixer whose trainable parameter count does not scale with the number of retained Fourier modes.

\paragraph{Test-set metrics.}
Table~\ref{tab:test_metrics_Cq5} summarizes the headline test-set performance under the same evaluation protocol as the classical baseline, including the VQC-based HQ-LP-FNO and the parameter-matched CM-LP-FNO control.

For temperature $\tilde{T}$, the VQC-based mixer lowers mean-error metrics (relative MAE from $2.9\%$ to $2.6\%$) but increases squared-error metrics (relative RMSE from $5.9\%$ to $6.7\%$). In contrast, the CM-LP-FNO control improves both typical and tail errors (relative MAE $2.3\%$ and relative RMSE $4.7\%$). This divergence suggests that the current VQC training 
configuration, which was not specifically optimized for 
squared-error objectives, produces occasional localized deviations near steep gradients and interfaces that disproportionately inflate RMSE, whereas the classical mode-shared mixer, trained under the same protocol, better suppresses such tails. Notably, we did not re-tune the training objective or loss weighting to explicitly optimize RMSE or overlap-based (IoU) metrics, targeted reweighting or interface-aware training may therefore further improve any of these variants.

For melt-related outputs, the difference is more pronounced. The VQC-based HQ-LP-FNO slightly degrades overlap quality and increases RMSE for both $\alpha$ and $f_l$ (e.g., IoU($f_l$) decreases from $0.911$ to $0.865$), whereas the parameter-matched CM-LP-FNO control improves both scalar errors and overlap quality, raising IoU($f_l$) to $0.970$ while also reducing the corresponding errors. Because CM-LP-FNO matches the VQC variant in parameter budget while changing only the mixer realization, this ablation isolates the source of the observed gains. The comparison indicates that mode-shared mixing is the key architectural principle, which the VQC naturally implements through its compact entanglement structure. While the classical control achieves lower RMSE and higher IoU on melt-related 
outputs, the VQC delivers the strongest mean temperature error among all three models (Table~\ref{tab:test_metrics_Cq5}), demonstrating that different mixer realizations can excel on complementary metrics. 

\begin{figure}[!htbp]
    \centering
    \includegraphics[width=\linewidth]{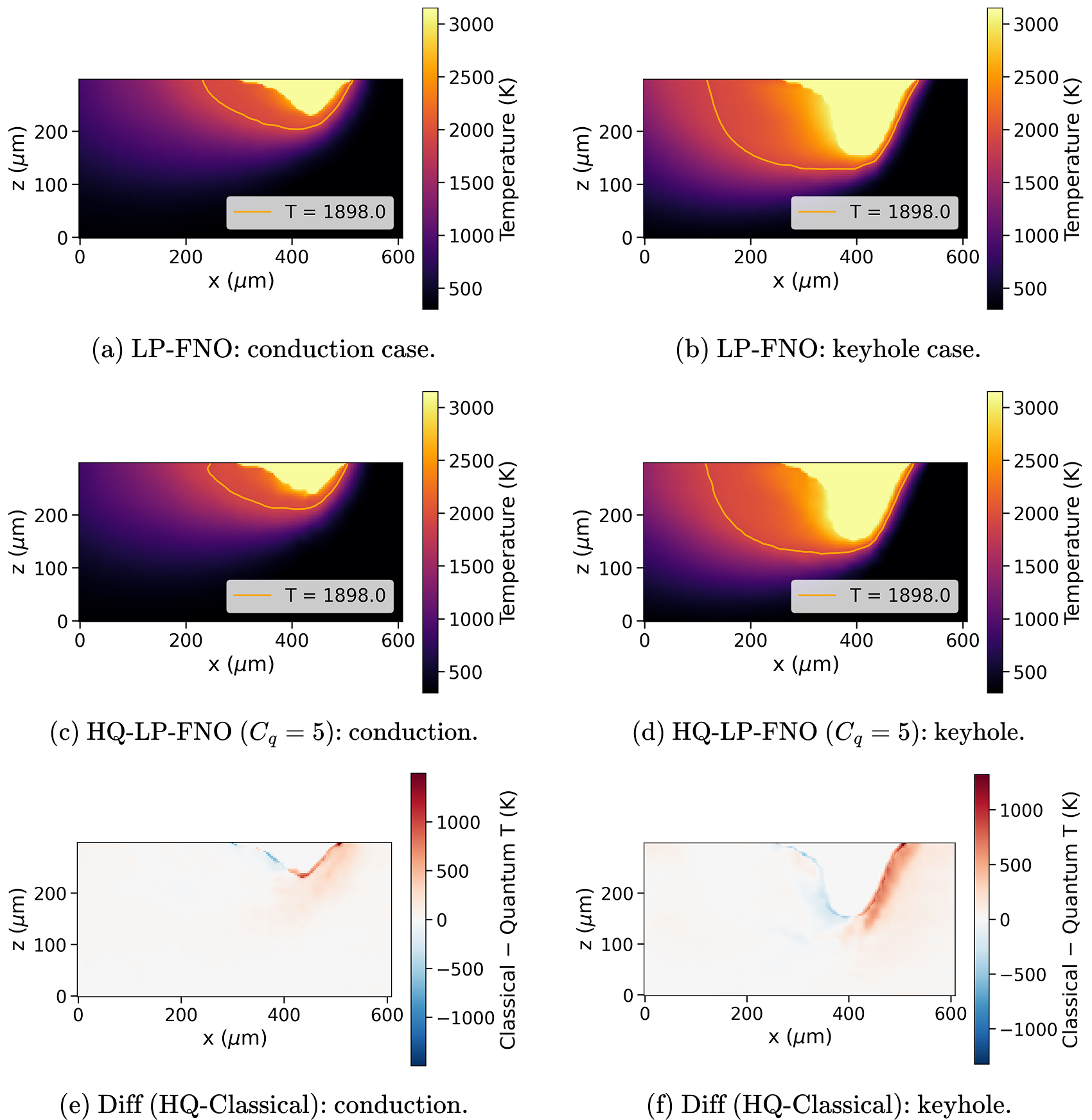}

    \caption{Qualitative comparison of classical and hybrid predictions at $\Cq=5$ for representative conduction and keyhole cases. The first two rows show temperature fields from each model using identical slice positions, colormaps, and color limits. The third row shows their difference on a diverging colormap centered at zero, where blue indicates lower temperature for the hybrid model and red indicates higher.}
    \label{fig:qualitative_cases_Cq5}
\end{figure}

\paragraph{Representative qualitative comparison.}
Fig.~\ref{fig:qualitative_cases_Cq5} compares classical LP-FNO and the VQC-based HQ-LP-FNO predictions on a representative conduction case and a keyhole case (the CM-LP-FNO control is omitted for brevity, see Table~\ref{tab:test_metrics_Cq5}). Across both examples, the differences are spatially localized: the bulk thermal field is similar, while the largest deviations occur near steep gradients and interfaces (notably around the melt-pool boundary and the keyhole wall). This localization is consistent with the mixed behavior in Table~\ref{tab:test_metrics_Cq5}: the hybrid model improves mean temperature error while the elevated RMSE is attributable to a small number of localized pointwise deviations near interfaces, regions where all models show the highest uncertainty and where interface-aware training objectives could yield further improvement.

\subsection{Process-space localization and interpretation}
\label{sec:results_localization}
To interpret Table~\ref{tab:test_metrics_Cq5} beyond a single scalar aggregate, we follow \cite{benoit2026lpfno} and localize errors over the process window in both $(P,\Vscan)$ and $(P,H^*)$ coordinates. Hybrid plots use the same axis definitions and color limits as the classical baseline, difference maps use a diverging colormap centered at zero.

\begin{figure}[!htbp]
    \centering
    \includegraphics[width=\linewidth,height=0.85\textheight,keepaspectratio]{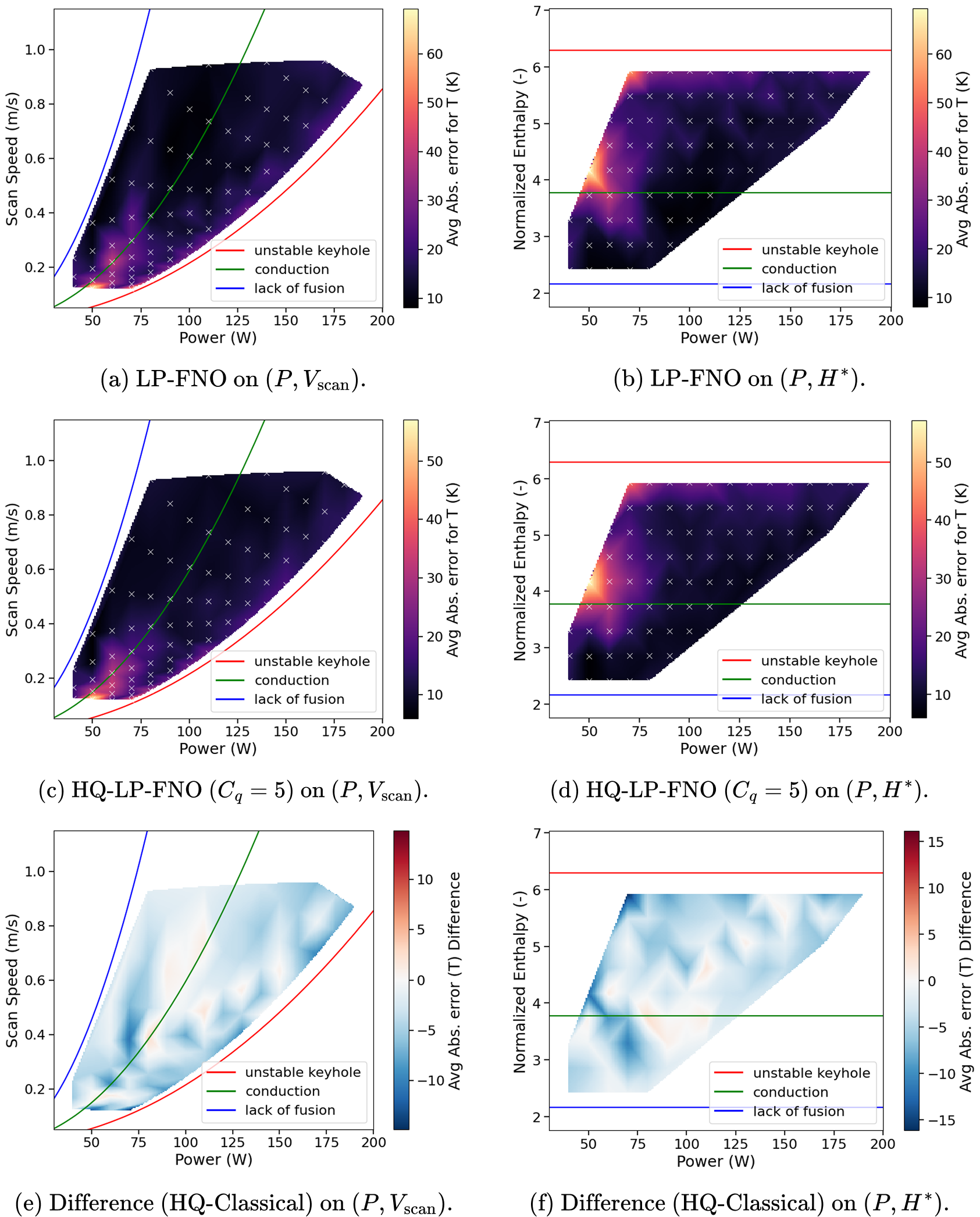}

    \caption{Temperature error maps across process space at $\Cq=5$.
    Rows show the classical model, the hybrid model, and their difference. Columns correspond to the $(P,\Vscan)$ and $(P,H^*)$ parameter planes. In the difference maps, blue indicates lower error for the hybrid model and red indicates higher error.}
    \label{fig:process_map_rmse_T}
\end{figure}

\paragraph{Process-space error maps.}
Figure~\ref{fig:process_map_rmse_T} shows average absolute temperature error (in K) over the sampled process window as a function of $(P,\Vscan)$ and $(P,H^*)$. The first two rows display classical and hybrid errors, the third row shows their pointwise difference (hybrid$-$classical). The difference maps are predominantly negative (blue) across most of the sampled window, indicating lower temperature MAE for the hybrid model at many parameter points. Visually, the magnitude of the reduction tends to be larger toward higher $P$ and higher $H^*$.

\paragraph{Interpretation.}
The VQC-based hybrid achieves its largest MAE reductions in the transition-to-keyhole regime (high $H^*$), where the classical baseline exhibits its highest errors. This suggests that the mode-shared VQC mixer provides a regularizing effect that is particularly beneficial in the most nonlinear region of the process space, where dense per-mode mixing may overfit to training configurations. Analogous error maps for $\alpha$ and 
$f_l$ are provided in Appendix~\ref{app:abs_error_maps}.

The broad MAE improvements coexist with elevated RMSE and lower IoU($f_l$) (Table~\ref{tab:test_metrics_Cq5}), reflecting a small number of localized pointwise deviations near steep gradients and phase interfaces
(Fig.~\ref{fig:qualitative_cases_Cq5}). This metric-dependent behavior indicates that the VQC mixer excels at reducing typical prediction error across the process space while remaining sensitive to sharp interface features under the current $L_1$-dominated training objective. Among the three output fields, $f_l$ is the most affected, as its thin spatial support amplifies the impact of localized deviations on both RMSE and IoU. These observations motivate interface-aware loss designs or multi-objective training strategies as a natural next step for the hybrid architecture.

\subsection{Quantum-specific analysis}
\label{sec:results_ratio_sweep}
The quantum-channel width $\Cq$ provides a continuous architectural knob that controls the balance between dense classical spectral mixing and compact VQC-based mixing (Sec.~\ref{sec:hybrid_lp_fno}). Characterizing the quantum design knob $\Cq$ remains valuable for understanding the mode-shared mixing principle and for informing future circuit designs \cite{liu2021rigorous}. The quantum-channel width $\Cq$ controls how many spectral output channels per retained Fourier mode are generated by the mode-shared VQC (Sec.~\ref{sec:hybrid_lp_fno}). By construction, increasing $\Cq$ allocates a larger share of the low-frequency channel-mixing outputs to the shared quantum mixer and reduces the size of the remaining classical dense spectral mixing block, yielding an explicit parameter-accuracy trade-off at fixed backbone, data, and training protocol. We benchmark a sweep over $\Cq\in\{0,3,5,8\}$ to quantify this trade-off, using the classical LP-FNO as the $\Cq=0$ reference.

\paragraph{Evaluation protocol and reporting (3 folds).}
For each $\Cq$ we keep all other settings fixed and evaluate on the same held-out test set using the primary metrics reported in Table~\ref{tab:test_metrics_Cq5}: relative mean error and relative RMSE on the masked temperature field $\tilde{T}$, and IoU($f_l$) for melt-pool overlap quality. In contrast to the eight-fold evaluation used elsewhere in this paper, the quantum sweep is summarized over \emph{three} folds due to the higher training cost of the hybrid models. Metrics are therefore reported as mean $\pm$ standard deviation across three folds.

\begin{table}[!htbp]
\centering
\caption{Quantum-channel-budget sweep (mean $\pm$ std across 3 folds).
IoU($\alpha$) is omitted as it exceeds $0.999$ for all settings.}
\label{tab:quantum_cq_sweep}
\renewcommand{\arraystretch}{1.15}
\setlength{\tabcolsep}{5pt}
\small
\begin{tabular}{@{}l cccc@{}}
\toprule
& $\Cq=0$ & $\Cq=3$ & $\Cq=5$ & $\Cq=8$ \\
& (classical) & & & \\
\midrule
Params (M)
  & 184.327 & 167.047 & 155.527 & \textbf{138.248} \\
Rel.\ mean $\tilde{T}$ [\%]
  & $2.53 \pm 0.48$ & {\boldmath$2.40 \pm 0.50$} & $2.58 \pm 0.61$ & $2.73 \pm 0.52$ \\
Rel.\ RMSE $\tilde{T}$ [\%]
  & $5.45 \pm 1.02$ & {\boldmath$5.31 \pm 1.08$} & $5.56 \pm 1.13$ & $5.61 \pm 1.07$ \\
IoU($f_l$)
  & {\boldmath$0.9047 \pm 0.0085$} & $0.9016 \pm 0.0079$ & $0.9003 \pm 0.0114$ & $0.90165 \pm 0.00868$ \\
\bottomrule
\end{tabular}
\end{table}

\paragraph{Parameter--accuracy trade-off and main observation.}
Table~\ref{tab:quantum_cq_sweep} shows a clear parameter reduction with stable primary metrics across the tested sweep range.
Increasing $\Cq$ from $0$ to $5$ reduces the trainable parameter count from $184.3$\,M to $155.5$\,M (about a $16\%$ reduction), while relative temperature errors and IoU($f_l$) remain within a narrow band: Rel.\ RMSE $\tilde{T}$ stays between $5.3$--$5.6\%$ and IoU($f_l$) varies only from $0.900$ to $0.905$, with all differences well within the inter-fold standard deviation.
Within this sweep, the trade-off is smooth: there is no evidence of a sharp threshold in Rel.\ RMSE $\tilde{T}$ or IoU($f_l$) over the tested $\Cq$ values.

\paragraph{Optimal quantum-channel allocation.}
Notably, $\Cq=3$ achieves the best temperature metrics across all tested configurations, including the fully classical baseline, with a relative MAE of $2.40\%$ and relative RMSE of $5.31\%$, while reducing the parameter count by $9.4\%$. This suggests that a moderate quantum-channel allocation, where the VQC handles a limited fraction of spectral channels, can improve predictive accuracy beyond what either fully classical dense mixing 
or larger quantum fractions achieve. The result points toward an optimal classical-quantum partitioning that balances the expressive capacity of the VQC mixer with the regularization induced by reducing dense over-parameterized blocks.

\paragraph{Interpretation: parameter-efficient mixing vs.\ over-parameterized dense blocks.}
This behavior is consistent with the hypothesis that the classical dense spectral channel-mixing block may be parameter-inefficient for its role in the overall LP-FNO pipeline: replacing a portion of dense, mode-wise mixing by a structured, mode-shared mixer can remove redundant degrees of freedom without degrading test-set generalization \cite{kontolati2023overparametrization, xiao2024amfno, tran2023ffno}.
In our hybrid layer, the VQC implements a constrained nonlinear channel mixer shared across Fourier modes, while the remaining backbone (3D FFT structure, retained modes, lifting/projection, and spatial decoding) is unchanged.
The observation that $\Cq=3$ improves over both the classical baseline and larger quantum fractions further suggests that the interplay between classical and quantum mixing channels admits a non-trivial optimum, motivating principled search strategies such as architecture-aware hyperparameter optimization for the hybrid partitioning.

\paragraph{Noise robustness.}
While the above characterizes the VQC in noiseless simulation, hardware execution requires tolerance to gate errors, decoherence, 
and finite-shot sampling. A noisy-simulator validation using backend-calibrated noise from \texttt{ibm\_torino} is presented in Appendix~\ref{app:noisy_simulator}, the quantum mixer remains numerically stable across the tested shot range.

\section{Conclusion}
\label{sec:conclusions}

We introduced HQ-LP-FNO, a hybrid quantum--classical neural operator that integrates a VQC into the spectral mixing pathway of a three-dimensional FNO for laser material interaction surrogates. The VQC acts as a mode-shared spectral mixer to generate expressive channel interactions while decoupling parameter growth from the retained Fourier mode budget.

Evaluated on the FLOW-3D WELD single-track Ti--6Al--4V dataset from~\cite{benoit2026lpfno}, HQ-LP-FNO achieves a $15.6\%$ reduction in trainable parameters compared to the classical LP-FNO baseline while improving mean-error metrics: relative MAE for temperature decreases from $2.89\%$ to $2.56\%$, and phase-fraction MAE is reduced by $26\%$. A sweep over the quantum-channel budget $\Cq$ reveals that a moderate allocation ($\Cq=3$) yields the best temperature metrics across all tested configurations, including the fully classical baseline, suggesting a non-trivial optimal partitioning between classical and quantum mixing channels. These results demonstrate that the proposed hybrid architecture achieves parameter-efficient, accurate surrogate modeling for complex multiphysics problems, with the mode-shared mixing principle.

To provide a rigorous evaluation framework, we co-introduced a parameter-matched classical bottleneck control (CM-LP-FNO), trained under identical conditions. The ablation confirms that mode-shared mixing, naturally implemented by the VQC through its compact circuit structure, is the key architectural principle 
driving the observed parameter-accuracy improvements.

To assess robustness under realistic execution conditions, we evaluated the quantum spectral mixer on a backend-calibrated noisy simulator derived from \texttt{ibm\_torino} (Appendix~\ref{app:noisy_simulator}). The mixer output remains numerically stable across the tested shot range, with mean squared deviation decreasing monotonically up to approximately $5000$ shots. This validation confirms that the VQC design tolerates gate errors, decoherence, and finite-shot sampling at the circuit depth considered, and provides concrete circuit-execution benchmarks for near-term hardware deployment.

Several directions remain open. Alternative circuit Ansatz, such as hardware-efficient or problem-inspired designs, may better exploit quantum correlations for spectral mixing~\cite{patapovich2025superposed}. Training objectives that explicitly penalize tail errors could improve the VQC's performance on geometry-sensitive metrics like IoU. Finally, evaluation on quantum hardware will be necessary to assess whether any representational benefits survive noise and limited connectivity.

\section*{Statements and Declarations}

\textbf{Competing interests.} The authors declare that they have no competing interests.

\textbf{Data availability.} The data supporting the findings of this study are not publicly available because they were generated using a licensed commercial simulation environment and are subject to project-specific and licensing restrictions. Data may be made available from the corresponding author upon reasonable request and subject to approval by the relevant project partners.

\textbf{Code availability.} The code used to train and evaluate the models is not publicly available at this time because it contains research software developed within an internal proprietary framework. Additional implementation details may be provided by the corresponding author upon reasonable request, subject to institutional and intellectual-property restrictions.

\textbf{Acknowledgements.} The authors would like to thank Terra Quantum AG and EMPA for supporting this work.

\bibliographystyle{apsrev}
\bibliography{references}

\newpage
\appendix

\section{Quantum Circuit Analysis}\label{sec:qc_analysis}

This section analyzes the VQC used as the
mode-shared spectral mixer. We report three complementary diagnostics:
\begin{itemize}
    \item redundancy analysis via ZX calculus,
    \item trainability analysis via Fisher information, and
    \item expressivity analysis via Fourier-series characterization.
\end{itemize}

\begin{figure}[H]
    \centering
    \includegraphics[width=\textwidth]{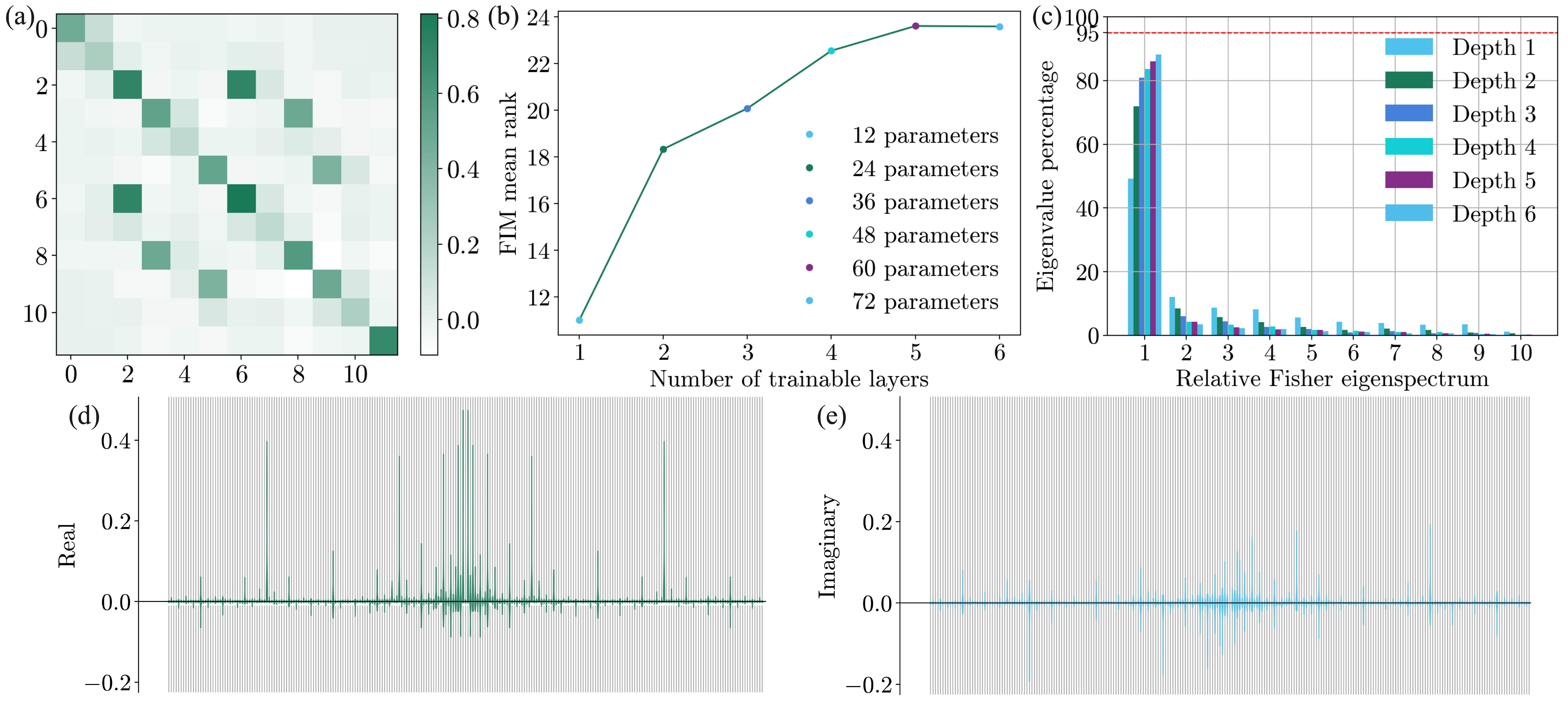}
    \caption{(a)~Average FIM showing diagonal dominance. (b)~FIM rank vs.\ number of trainable layers. (c)~Eigenvalue distribution across depths. (d--e)~Fourier coefficient spectrum, real and imaginary parts.}
    \label{fig:QC analysis}
\end{figure}

\subsection{Redundancy analysis via ZX calculus}\label{sec:appendix_ZX}

ZX calculus is a graphical language for representing and simplifying quantum circuits~\cite{coecke2008zx}. Within this formalism, a circuit is mapped to a graph comprising nodes (``spiders'') connected by edges. Rewrite rules are then applied to eliminate redundant elements~\cite{wetering2020zx}, and the reduced graph is compared with the original to quantify structural redundancy.

Redundancy is quantified by the fraction of initial trainable parameters preserved after simplification. A larger retained fraction indicates higher architectural quality, a circuit that admits no reduction is classified as ZX-irreducible.

The optimization procedure preserves all $12$ circuit parameters (100\%), indicating that the odd even IsingXY ansatz with QFT bookends contains no structurally detectable redundancy at this depth and qubit count.

The ZX-irreducibility confirms that the ansatz does not contain trivially canceling rotations or identity-equivalent gate sequences. However, ZX calculus tests only structural redundancy, it does not guarantee trainability or expressivity, which are assessed separately below.

\subsection{Trainability assessment via Fisher information}\label{sec:appendix_fim}

We assess trainability using the Fisher information matrix (FIM), which quantifies the sensitivity of the circuit output distribution to parameter changes~\cite{abbas2021powerofqnns, amari1998gradient, haboury2024information}. The FIM defines a Riemannian metric on the parameter manifold:
\begin{equation}
    F(\theta)=\mathbb{E}_{\{x_i,y_i\}}\!\left[\nabla_\theta \log P(y\mid x,\theta)\,\nabla_\theta \log P(y\mid x,\theta)^T\right].
\end{equation}

To analyze local sensitivity, the metric is diagonalized to obtain a locally Euclidean tangential basis, the diagonal entries correspond to squared gradient magnitudes in this basis and are given by the eigenvalues of the FIM \cite{holmes2022connecting, larocca2023overparam}. This analysis is relevant for diagnosing and mitigating the barren-plateau phenomenon, which manifests as vanishing gradients in a quantum neural network with increasing qubit count.

As shown in Ref.~\cite{mcclean2018barren}, gradient expectation values approach zero and gradient variances diminish exponentially with the number of qubits. This behaviour occurs when gradients concentrate near zero, so that a substantial fraction of parameters do not effectively participate in training. Accordingly, inspection of the FIM eigenvalue spectrum across multiple instances of \( \theta \) provides a diagnostic for trainability and resilience against barren plateaus. In particular, improved trainability is associated with reduced eigenvalue degeneracy.

Following Ref.~\cite{abbas2021powerofqnns}, a Gaussian-distributed dataset \( x \sim \mathcal{N}(\mu = 0, \sigma^2 = 1) \) is generated. The joint probability is then obtained by evaluating the overlap between the prepared quantum state and the state produced by the quantum layer:
\begin{equation}
    P(y, x\mid\theta) = |\braket{y|\psi (\theta, x)}|^2
\end{equation}
Here, \(y\) denotes the output state. Averaging over \(x\) yields the Fisher information for any given \(\theta\).

The result of this analysis includes multiple plots. In Fig.~\ref{fig:QC analysis}(a), the average FIM exhibits a pronounced diagonal structure, indicating that a large fraction of parameters are active and contribute to the performance of the circuit. The near absence of off-diagonal entries indicates that no pair of parameters is sufficiently coupled to justify significant structural changes. In Fig.~\ref{fig:QC analysis}(c), the eigenvalue distribution displays multiple parameters with eigenvalues substantially above zero. Approximately $50\%$ of parameters exhibit near-zero eigenvalues, while still remaining far below the $95\%$ threshold. Most remaining parameters retain appreciable influence, show a closer-to-uniform distribution (relative to other depths), and contribute to trainability.

\subsection{Expressivity assessment via Fourier series}\label{sec:appendix_fourier}

Ref.~\cite{schuld2021encoding} shows that the output of a parameterized quantum circuit may be expressed as a truncated Fourier series. For a feature vector \(x\) of length \(N\), this representation takes the form
\begin{equation}
    f_\theta (x) = \sum_{\omega_1 \in \Omega_1} \cdots \sum_{\omega_N \in \Omega_N} c_{\omega_1,\ldots,\omega_N}(\theta)\, e^{-i\,\omega\cdot x},
\end{equation}
where each frequency component satisfies \(\omega_k \in \{-d,\ldots,0,\ldots,d\}\).

Equivalently, the number of Fourier terms is given by one plus twice the number of times the input is embedded in the circuit, denoted by \(d\). In this analysis, expressivity of \( f_{\theta}(x) \) is assessed by sampling \(\theta_i\) uniformly at random from \([0,2\pi]\) and selecting equidistant \(x\) values with sampling frequency \(d\).

Two aspects are emphasized: the number of accessible terms (degree) and the accessibility of coefficients associated with each term (coefficient expressivity). Expressivity is quantified by the number of nonzero coefficients in the Fourier spectrum, a larger number indicates greater expressivity.

Our VQC exhibits a high degree of expressivity, with $485$ nonzero coefficients out of $485$ admissible terms (i.e., $100\%$), as shown in Fig.~\ref{fig:QC analysis}(d--e). This indicates high expressivity of the implemented model, however, it should be emphasized that many amplitudes are small and may increase with additional layers (subject to the available computational budget).

\subsection{Summary}

The quantum circuit implemented in this paper exhibits a low degree of redundancy and a high degree of trainability, with a majority of parameters remaining active and contributing substantially to performance. Since a non-negligible fraction of near-zero eigenvalues is observed (suggesting potential susceptibility to barren-plateau effects under further scaling), the model is chosen to remain in an underparameterized regime. Concurrently, expressivity demonstrates a $100\%$ nonzero-coefficient rate, with the possibility of increasing coefficient magnitudes further through the addition of new layers.

\section{Noisy-simulator validation of the quantum spectral mixer}
\label{app:noisy_simulator}

\begin{figure}[H]
  \centering
  \includegraphics[width=0.8\linewidth]{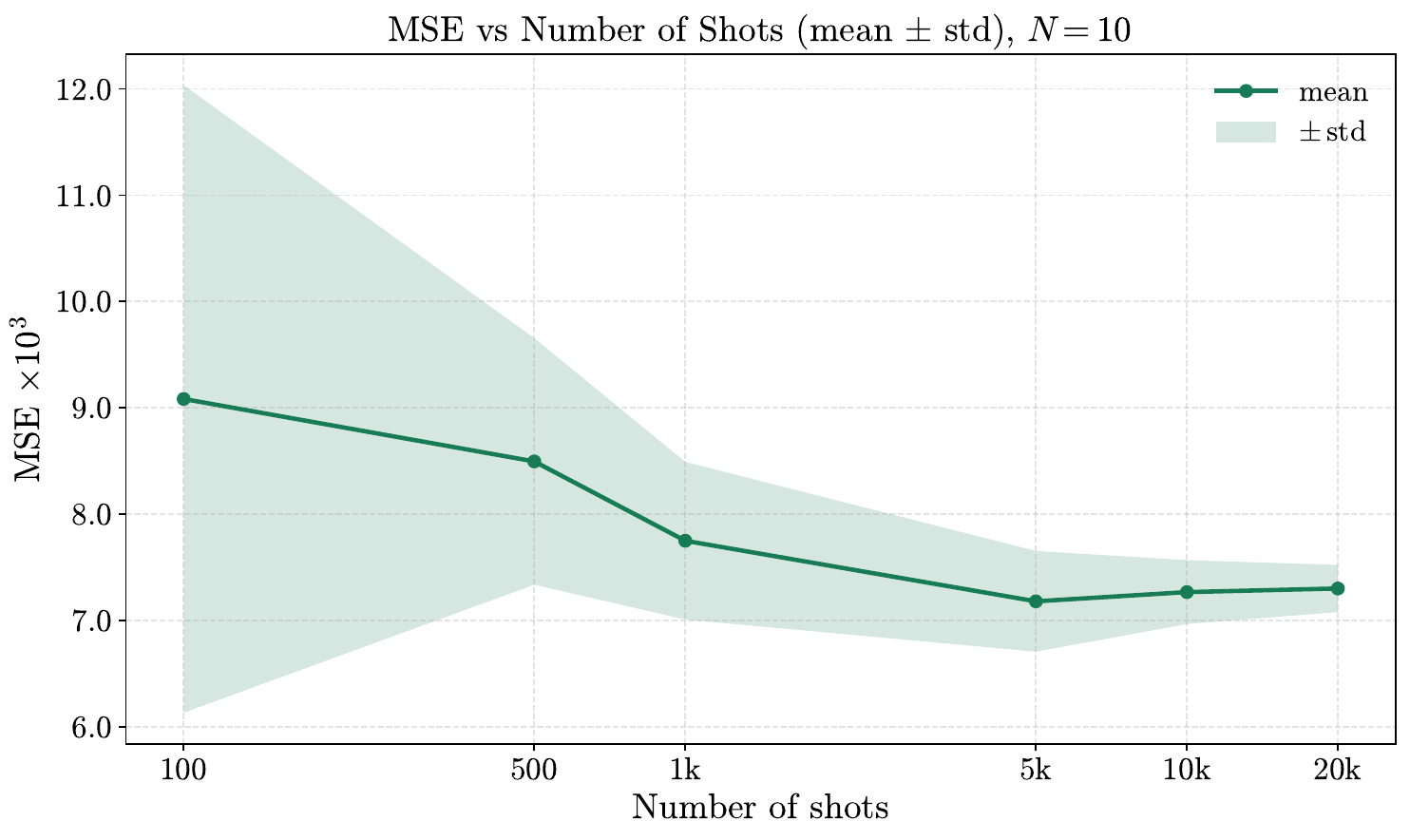}
  \caption{Mean squared error between the noiseless and noisy outputs of the quantum spectral mixer as a function of the number of measurement shots under the \texttt{ibm\_torino}-derived noise model. Points show the mean over 10 repeated runs, the shaded band indicates $\pm$ one standard deviation.}
  \label{fig:mse_shots}
\end{figure}

To bridge the gap between idealized (noise-free) circuit simulation and hardware execution, we evaluate the quantum spectral mixer under a backend-calibrated noisy simulator corresponding to \texttt{ibm\_torino}, a quantum device operated by IBM Quantum (Heron r1 family)~\cite{ibm-quantum-platform}. While the main results in Sec.~\ref{sec:results} are obtained with noiseless simulation to isolate representational effects, device-level execution is unavoidably affected by decoherence, gate errors, readout noise, and finite-shot sampling~\cite{preskill2018nisq, kuzmin2025method}.

The experiment uses the hybrid configuration from Sec.~\ref{sec:results} ($C_q = 5$, $n_q = 5$, depth $d = 1$), without retraining, only the inference stage of the quantum branch is evaluated under noisy execution. A noise model is constructed from \texttt{ibm\_torino} backend calibration data, including native basis gates, coupling map, gate error rates, relaxation times, and readout assignment errors. Circuits are executed with a finite number of measurement shots to estimate the Pauli-$Z$ expectation values required by the quantum mixer. This workflow is implemented using Qiskit~\cite{qiskit2024}.

To isolate the effect of finite-shot sampling and backend-derived noise, we probe the quantum spectral mixer in isolation. For a fixed representative retained Fourier-mode input, we compare the shot-based noisy-simulator output to a noiseless reference. Figure~\ref{fig:mse_shots} reports the mean squared deviation versus the number of shots (mean $\pm$ std over 10 runs with different sampling seeds).

The resulting curve shows a monotonic decrease of the mean squared deviation with increasing shot count, consistent with the expected reduction of sampling-induced fluctuations. Beyond approximately $5000$ shots, the curve approaches a plateau, indicating diminishing returns from further shot increases at the circuit depth considered. Importantly, the mixer output remains numerically well-behaved across the explored range, with no instability or divergence observed.

From a resource perspective, it is instructive to estimate the number of circuit executions per sample. With three Fourier layers ($L = 3$) and retained modes $(m_x, m_y, m_z) = (25, 20, 15)$, and accounting for the four-corner layout in $(k_x, k_y)$, the architecture requires $L \times 4 \times m_x m_y m_z \approx 9 \times 10^4$ circuit evaluations per forward pass. Multiplying by the shot count $S$ yields roughly $(9 \times 10^4) \times S$ total executions per sample, which quickly becomes the dominant cost driver for end-to-end evaluation.

The circuit-execution budget can be reduced by decreasing the retained-mode count, enabling the quantum mixer in a subset of Fourier layers, or restricting it to the lowest-frequency modes while keeping the remainder classical. Additionally, adaptive shot allocation and readout-error mitigation~\cite{cai2023quantum} can lower the per-mode shot requirement. This validation provides both a numerical robustness check and a concrete resource estimate informative for future hardware experiments.

\newpage

\section{FNO Hyperparameters}
\label{app:fno_hparams}

This section summarizes the architecture and training hyperparameters used throughout the experiments. The table reports the settings of the FNO backbone, decoder, loss aggregation scheme, and the optimization setup.

\begin{table}[!htbp]
\centering
\caption{FNO architecture and training hyperparameters. We follow \cite{benoit2026lpfno}; for hybrid models we use a cosine-annealing schedule with a higher base learning rate.}
\label{tab:fno_hparams}
\small
\setlength{\tabcolsep}{4pt}
\begin{tabular}{lll}
\toprule
Category & Parameter & Value \\
\midrule
\multicolumn{3}{l}{\textit{FNO backbone}}\\
 & Spatial dimension & 3 \\
 & Number of FNO layers ($L$) & 3 \\
 & Latent channel width ($C$) & 32 \\
 & Retained Fourier modes $(m_x,m_y,m_z)$ & $[25,\,20,\,15]$ \\
 & Padding & 9 (constant) \\
 & Activation function & GELU \\
 & Coordinate features & enabled \\
 & Input variables & $x,\,y,\,z,\,\Vscan,\,P$ \\
\midrule
\multicolumn{3}{l}{\textit{Decoder (pointwise network)}}\\
 & Architecture & convolutional fully connected \\
 & Number of layers & 3 \\
 & Hidden layer width & 32 \\
 & Activation function & SiLU \\
 & Weight normalization & enabled \\
 & Skip connections & disabled \\
 & Output variable & $T$ \\
\midrule
\multicolumn{3}{l}{\textit{Loss aggregator (ReLoBRaLo \cite{bischof2025relobralo})}}\\
 & $\alpha$ & 0.95 \\
 & $\beta$ & 0.99 \\
 & $\tau$ & 3.0 \\
 & $\varepsilon$ & \num{1e-8} \\
\midrule
\multicolumn{3}{l}{\textit{Optimization}}\\
 & Optimizer & Lion \cite{chen2023lion} \\
 & Learning rate & \num{6e-5} (classical); \num{1e-4} (hybrid) \\
 & $\beta$ coefficients & $(0.9,\,0.99)$ \\
 & Weight decay & 0 \\
\midrule
\multicolumn{3}{l}{\textit{Learning-rate schedule}}\\
 & Scheduler & exp.\ decay (classical); cosine annealing (hybrid) \\
 & exp.\ decay: rate / steps & 0.98 / 100 \\
 & cosine: $T_{\max}$ / $\eta_{\min}$ & 6000 / \num{1e-5} \\
\midrule
\multicolumn{3}{l}{\textit{Training setup}}\\
 & Maximum training steps & 6000 \\
 & Gradient clipping norm & 0.5 \\
 & Batch size (training / validation) & 1 / 5 \\
\bottomrule
\end{tabular}
\end{table}

\FloatBarrier
\newpage

\section{Avg. Absolute Error Maps for $\alpha$ and $f_l$}
\label{app:abs_error_maps}

This section provides process-space maps of the average absolute error for $\alpha$ and $f_l$ at $\Cq=5$, complementing the temperature-based analysis discussed in the main text. For each target variable, the figures compare the classical LP-FNO, the hybrid HQ-LP-FNO, and their pointwise difference.

\begin{figure}[H]
    \centering
    \includegraphics[width=\linewidth,height=0.85\textheight,keepaspectratio]{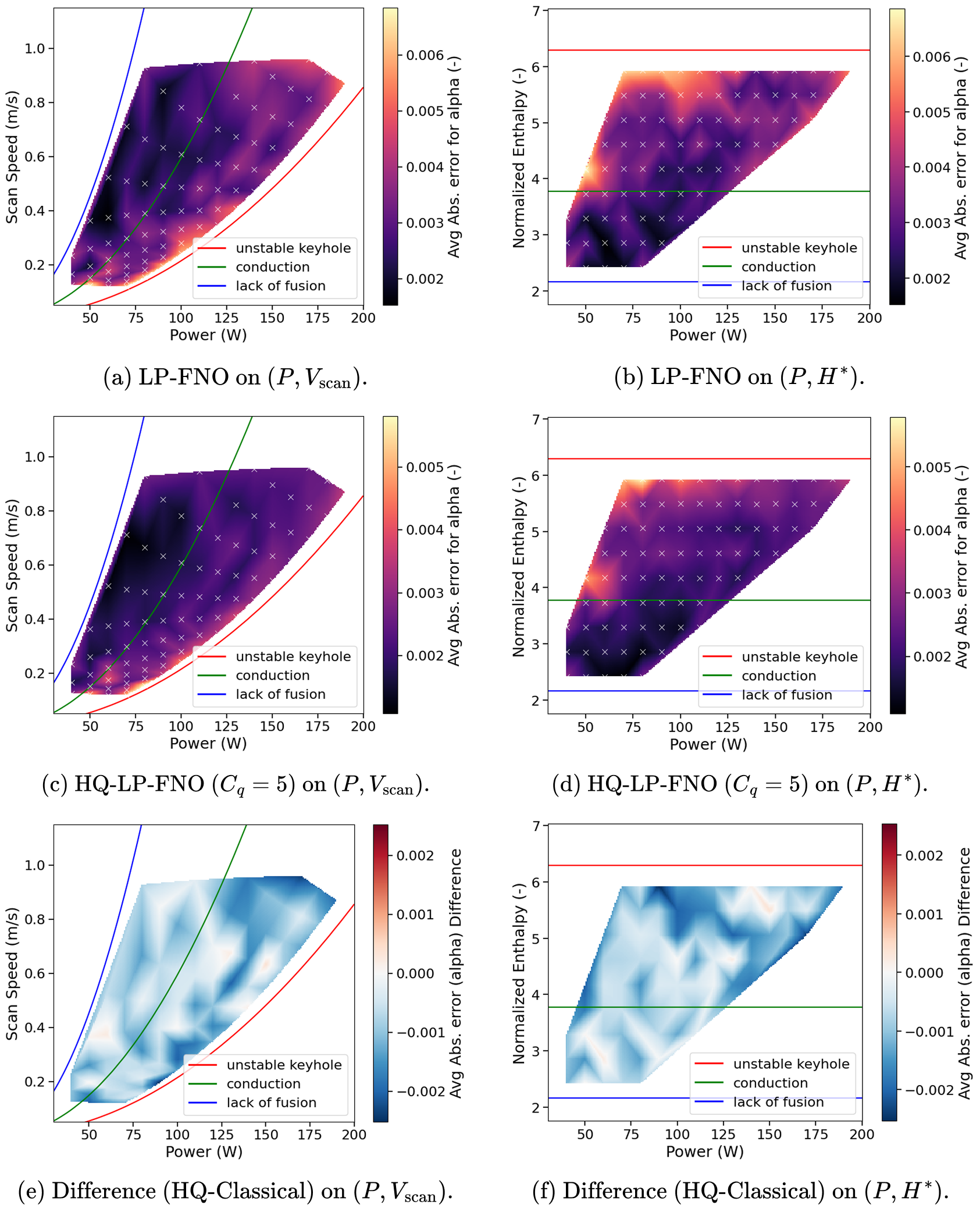}

    \caption{$\alpha$ error maps across process space at $\Cq=5$, analogous to Fig.~\ref{fig:process_map_rmse_T} for temperature.}
    \label{fig:process_map_rmse_alpha}
\end{figure}

\begin{figure}[H]
    \centering
    \includegraphics[width=\linewidth,height=0.85\textheight,keepaspectratio]{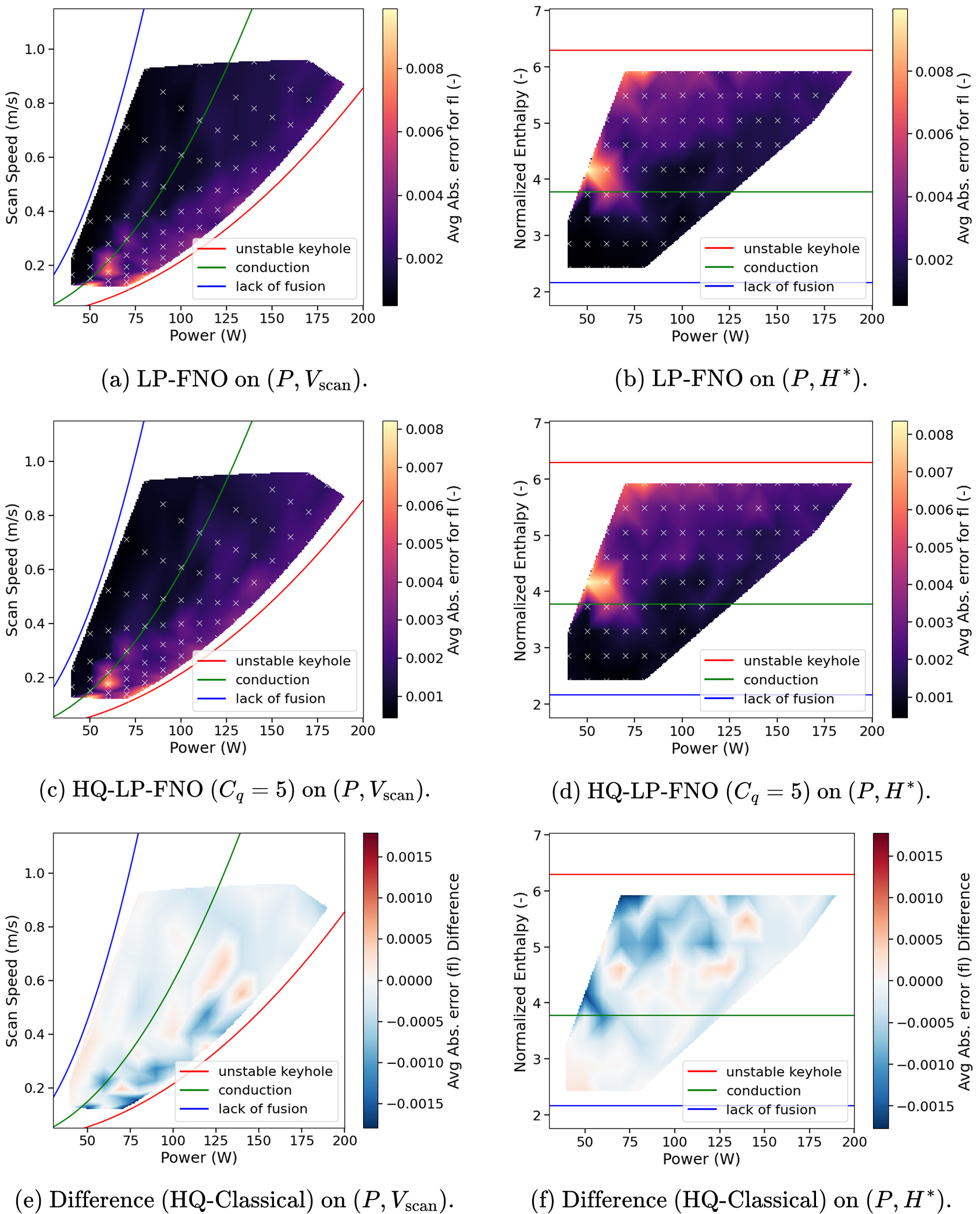}

    \caption{$f_l$ error maps across process space at $\Cq=5$, analogous to Fig.~\ref{fig:process_map_rmse_T} for temperature}
    \label{fig:process_map_rmse_fl}
\end{figure}

\FloatBarrier
\newpage

\section{IoU Error Maps for $\alpha$ and $f_l$}
\label{app:iou_error_maps}

This section reports process-space maps of the IoU error, defined as $1-\mathrm{IoU}$, for $\alpha$ and $f_l$ at $\Cq=5$. Using the same layout as in Appendix C, the figures compare the classical and hybrid models across the $(P,\Vscan)$ and $(P,H^*)$ parameter planes and visualize their difference.

\begin{figure}[H]
    \centering
    \includegraphics[width=\linewidth,height=0.85\textheight,keepaspectratio]{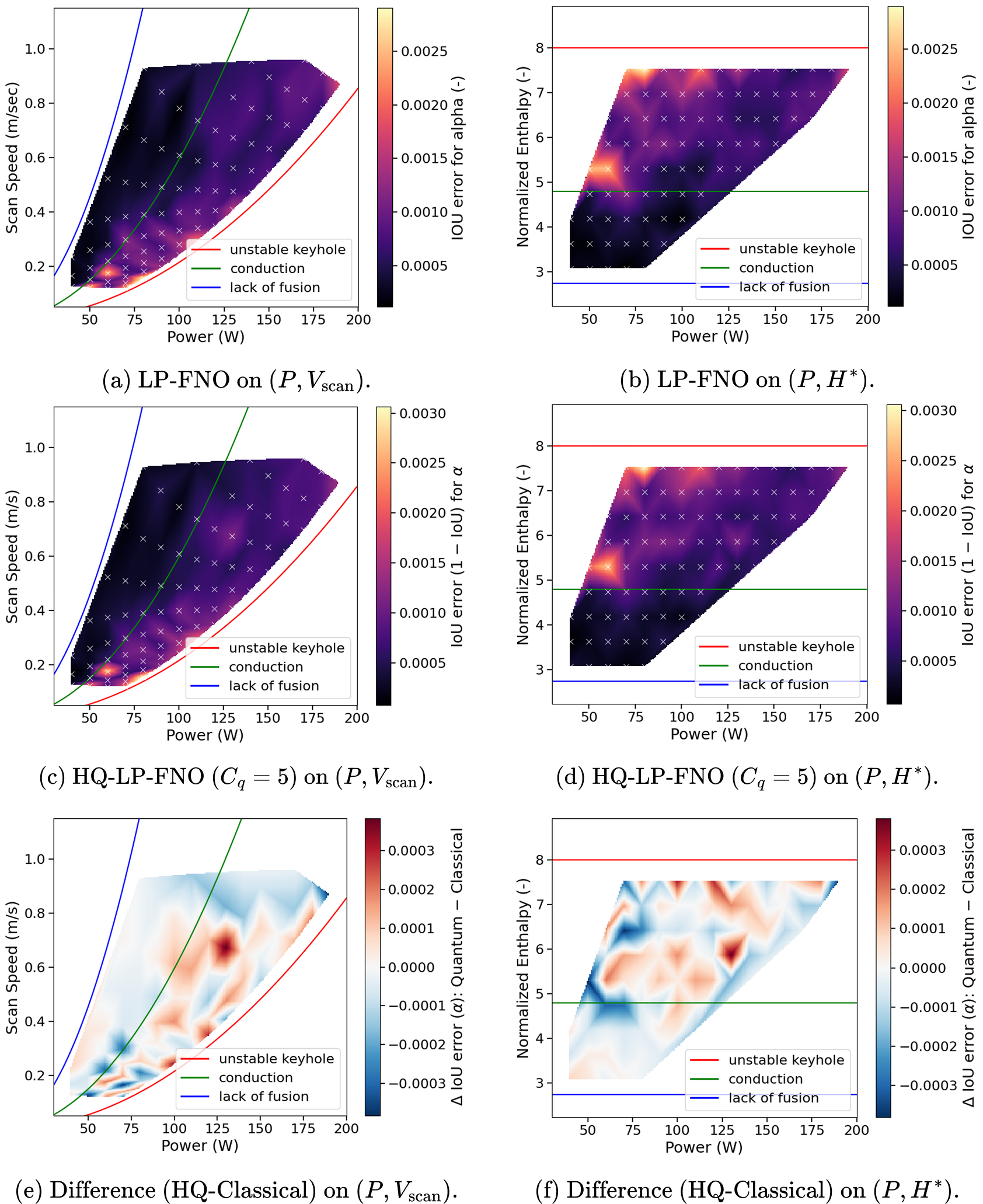}
    \caption{$\alpha$ IoU error ($1 - \text{IoU}$) maps across process space at $\Cq=5$.}
    \label{fig:process_map_iou_alpha}
\end{figure}

\begin{figure}[H]
    \centering
    \includegraphics[width=\linewidth,height=0.85\textheight,keepaspectratio]{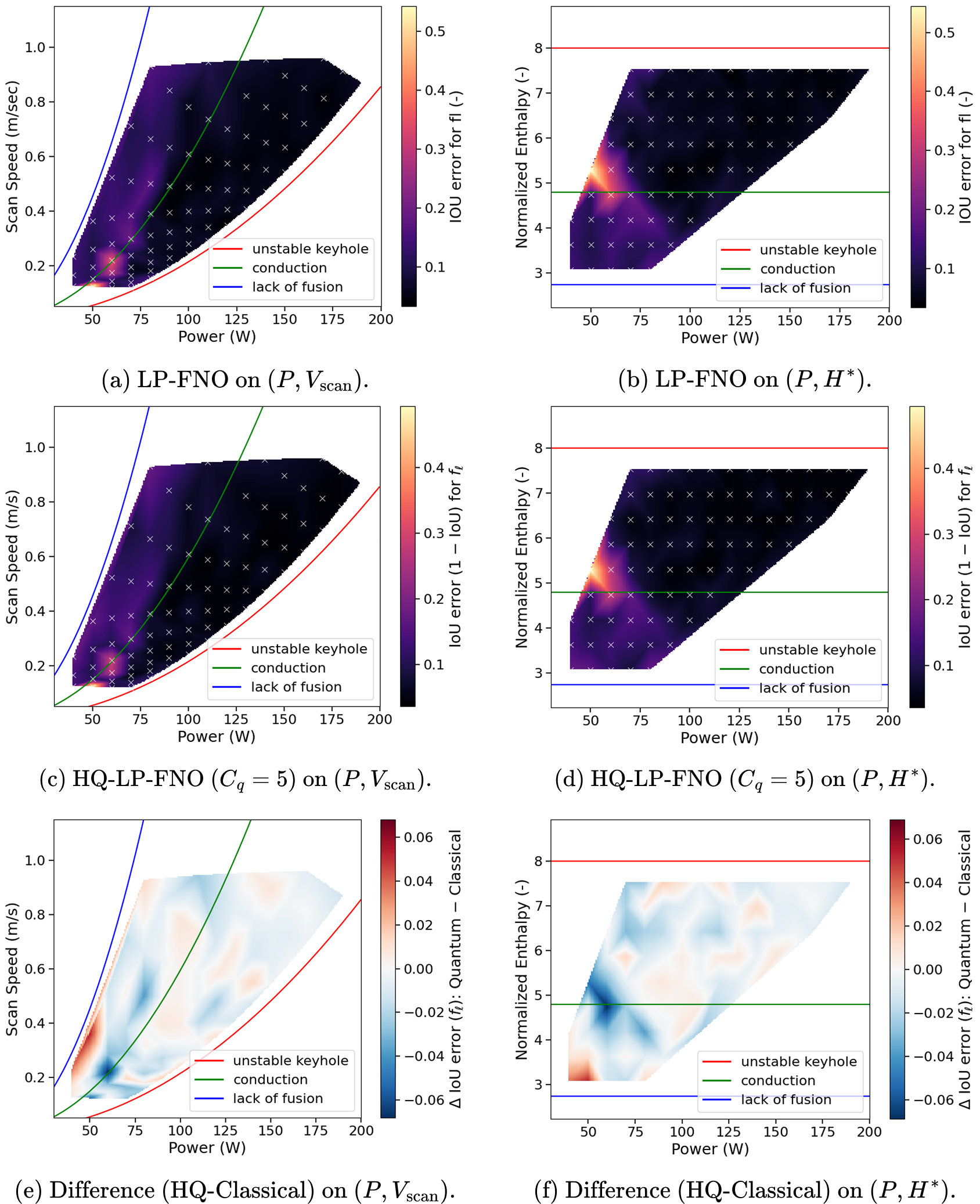}
    \caption{$f_l$ IoU error ($1 - \text{IoU}$) maps across process space at $\Cq=5$.}
    \label{fig:process_map_iou_fl}
\end{figure}

\FloatBarrier

\end{document}